\documentclass[11pt,a4paper]{article}
\pdfoutput=1
\usepackage[latin9]{inputenc}
\usepackage{xcolor}
\usepackage{longtable}
\usepackage{float}
\usepackage{rotfloat}
\usepackage{amsmath}
\usepackage{amssymb}
\usepackage{graphicx}
\usepackage[unicode=true,
 bookmarks=false,
 breaklinks=false,pdfborder={0 0 1},backref=false,colorlinks=true]
 {hyperref}
\hypersetup{
 linkcolor=blue,urlcolor=magenta,citecolor=blue}
\usepackage{fullpage}
\makeatletter

\providecommand{\tabularnewline}{\\}

\usepackage[margin=0.5cm]{caption}
\usepackage{cite}
\usepackage{upgreek}

\newcommand{\be}{\begin{equation}}
\newcommand{\ee}{\end{equation}}
\newcommand{\bea}{\begin{eqnarray}}
\newcommand{\eea}{\end{eqnarray}}
\newcommand{\bg}{\begin{gather}}
\newcommand{\eg}{\end{gather}}
\newcommand{\bseq}{\begin{subequations}}
\newcommand{\eseq}{\end{subequations}}

\renewcommand{\ln}{\mathop{\rm ln}\nolimits}

\newcommand{\ket}[1]{| #1 \rangle}

\usepackage{braket}

\definecolor{col3}{rgb}{1.0, 0.0, 1.0}
\definecolor{col4}{rgb}{0.625, 0.0, 0.0}

\providecommand*{\eu}%
{\ensuremath{\mathrm{e}}}
\providecommand*{\iu}%
{\ensuremath{\mathrm{i}}}

\providecommand*{\du}%
{\ensuremath{\mathrm{d}}}
\date{}

\@ifundefined{showcaptionsetup}{}{%
 \PassOptionsToPackage{caption=false}{subfig}}
\usepackage{subfig}
\makeatother

\begin{document}
\baselineskip=15.5pt
\begin{titlepage}
\begin{center}
{\Large\bf   Glueball Spins in $ D=3$ Yang-Mills }\\
\vspace{0.5cm}
{ \large
Peter Conkey$^{a}$, Sergei Dubovsky$^{a}$, and Michael Teper$^{b}$ 
}\\
\vspace{.45cm}
{\small  \textit{  $^a$Center for Cosmology and Particle Physics,\\ Department of Physics,
      New York University\\
      New York, NY, 10003, USA}}\\ 
      \vspace{.1cm}
      \vspace{.25cm}      
      {\small \textit {{$^b$Rudolf Peierls Centre for Theoretical
Physics,\\Clarendon Laboratory, University of Oxford,\\ Parks Road, Oxford OX1 3PU, UK\\
\centerline{and}
All Souls College, University of Oxford,\\
High Street, Oxford OX1 4AL, UK}}}
\end{center}
\begin{center}
\begin{abstract}
We determine spins of more than 100 low lying glueball states in $D=2+1$ dimensional $SU(4)$ gluodynamics by a lattice calculation. We go up to $J=8$ in the spin value.
We compare the resulting spectrum with predictions of the Axionic String Ansatz (ASA). We find a perfect match for 39 lightest states, corresponding to the first four string levels.
In particular, this resolves tensions between the ASA predictions and earlier spin determinations. The observed spins of heavier glueballs are also in a good agreement with the ASA. We did not identify 
any sharp tension between lattice data and the ASA, but more work is needed to fully test the ASA predictions for the spins of 64 states at the fifth string level.
\end{abstract}
\end{center}
\end{titlepage}

\newpage{}

\section{Introduction}
$SU(N_c)$ Yang--Mills (YM) theory turns into a free string theory in the 't Hooft planar limit $N_c\to\infty$ \cite{'tHooft:1973jz}.
Understanding these confining strings  remains an outstanding challenge. Lattice Monte-Carlo simulations deliver precious 
experimental guidance for this enterprise. 
Lattice  observables sensitive to the stringy origin of YM glueballs belong to two large distinct categories.
The most direct way to probe the worldsheet dynamics of confining strings on the lattice is to measure the spectrum of tolerons---long flux tubes wound around one of the spatial directions \cite{Athenodorou:2010cs,Athenodorou:2011rx,Athenodorou:2013ioa,Athenodorou:2016kpd,Athenodorou:2017cmw}.  Applying the effective string theory methods, such as large radius expansion 
\cite{Luscher:1980ac,Luscher:2004ib,Aharony:2010cx,Aharony:2010db,Dubovsky:2012sh,Aharony:2013ipa}, Thermodynamic Bethe Ansatz (TBA) \cite{Dubovsky:2013gi,Dubovsky:2014fma} or flux tube $S$-matrix bootstrap
\cite{EliasMiro:2019kyf} one can then extract  properties of the worldsheet action and $S$-matrix from the spectral data. These studies resulted in the Axionic String Ansatz (ASA) for the worldsheet theory in 
 both $D=3$ and $D=4$ gluodynamics \cite{Dubovsky:2015zey,Dubovsky:2016cog}.

The second set of lattice data comes from glueball spectroscopy \cite{Meyer:2002mk,Meyer:2003wx,Meyer:2004gx,Lucini:2010nv,Athenodorou:2016ebg}. Glueball spectra have an advantage of representing traditional physical observables, directly  measurable in the lab. However, relating  these data to the worldsheet properties is hard. In particular, effective string theory is only useful in the vicinity of the leading Regge trajectory \cite{Hellerman:2013kba}.
Hopefully, one can do better with the ASA, which postulates that the worldsheet theory can be treated as a deformation of an integrable model (see \cite{Dubovsky:2018vde,Donahue:2019adv} for a 
further discussion of the underlying physics). To predict the glueball spectrum one needs then to solve for the short string spectrum in the integrable approximation, and to develop perturbative 
methods accounting for violations of integrability.

A very first step towards implementation of  this program has been made in \cite{Dubovsky:2016cog} for $D=3$ YM. The ASA string is particularly simple at $D=3$, where it belongs to the same equivalence class as the minimal Nambu--Goto string with no additional massless or massive worldsheet degrees of freedom present. In the integrable approximation the ASA worldsheet $S$-matrix is the same as the worldsheet $S$-matrix of
critical strings \cite{Dubovsky:2012wk}. Building on this similarity, \cite{Dubovsky:2016cog} came up with a prediction of glueball quantum numbers. At $D=3$ these are glueball spin $J$, spatial parity $P$ and 
charge conjugation parity $C$.  $J>0$ glueball states come out as doublets of states carrying opposite spatial parities $P$, so that $P$ carries non-trivial information only for $J=0$ glueballs. We will refer to the resulting spectrum as the ASA spectrum, but it is important to keep in mind that some additional assumptions have been made in  \cite{Dubovsky:2016cog}, which may be not a part of the ASA.  

In the present paper we report  results of a dedicated lattice simulation whose goal is to test the predictions of \cite{Dubovsky:2016cog}. The major difficulty for such a measurement is that the physical $SO(2)$ rotation symmetry is broken to its $\mathbb{Z}_4$ subgroup on a cubic lattice. As a result, the conventional glueball mass measurements, such as the ones presented in \cite{Athenodorou:2016ebg}, deliver only  $\mathbb{Z}_4$ transformation properties of the glueball states, rather than the physical glueball spins.

A technique to measure the actual physical spins $J$ was described in \cite{Meyer:2002mk,Meyer:2003wx,Meyer:2004gx}, where it was applied to several low lying glueball states. Here we build on this proposal and perform continuous spin measurements for all accessible glueball states (in practice, this is more than 100 states). Let us stress that our main goal is to provide spin determinations rather than to perform a high precision mass measurement. As a result, the mass values which came out from our simulation are less precise than those reported in \cite{Athenodorou:2016ebg}. Also we restrict ourselves to the $SU(4)$ gauge group. This value of $N_c$ should be large enough for our purposes, given that the typical mass shifts between $N_c=4$ and $N_c=\infty$ measured in \cite{Athenodorou:2016ebg} are all within 10$\%$.

\begin{figure}[H]
\begin{center}
\includegraphics[width=1\textwidth]{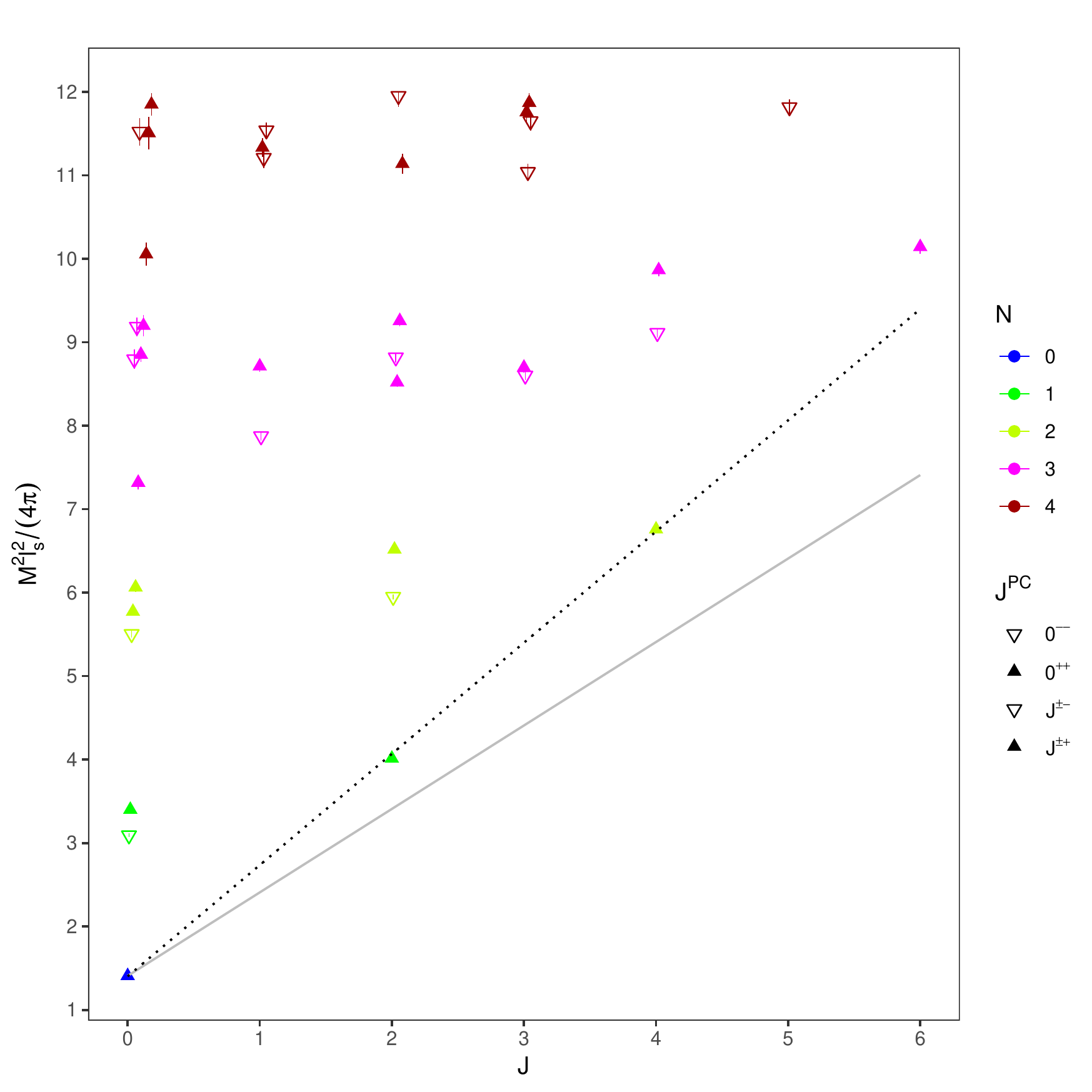}

\caption{The Chew-Frautschi plot  with color coded level  assignment up
  to $N=3$. The dotted line is the weighted least squares fit to leading Regge trajectory
$M^{2}l_{s}^{2}=4\pi\left(1.40(4)+1.35(4)J\right)$. The light solid line
is the linear Regge trajectory with a unit slope line starting at $0^{++}$ ground state.
\label{fig:Glueball-spectrum-byN-toNequals3}}
\end{center}
\end{figure}

\begin{figure}[H]
\begin{center}
\includegraphics[width=1.0\textwidth]{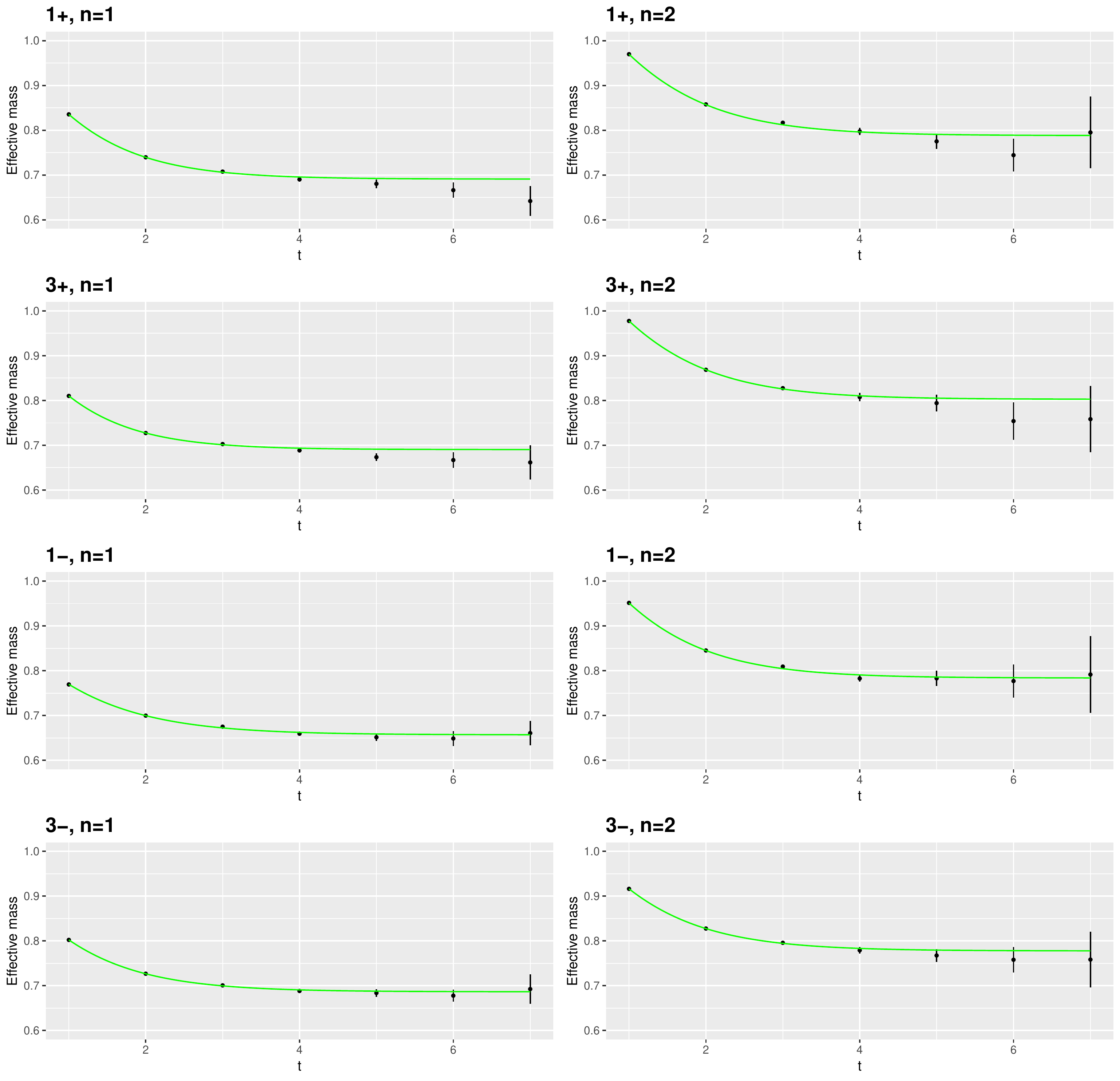}

\caption{Plot of effective mass $E_{eff}(t)$ versus $t$ (with fit, green curve)
for  lowest energy $J^{C}=1^{\pm},3^{\pm}$ states.\label{fig:MassplotExample}}
\end{center}
\end{figure}

\begin{figure}[H]
\begin{center}
\includegraphics[width=1.0\textwidth]{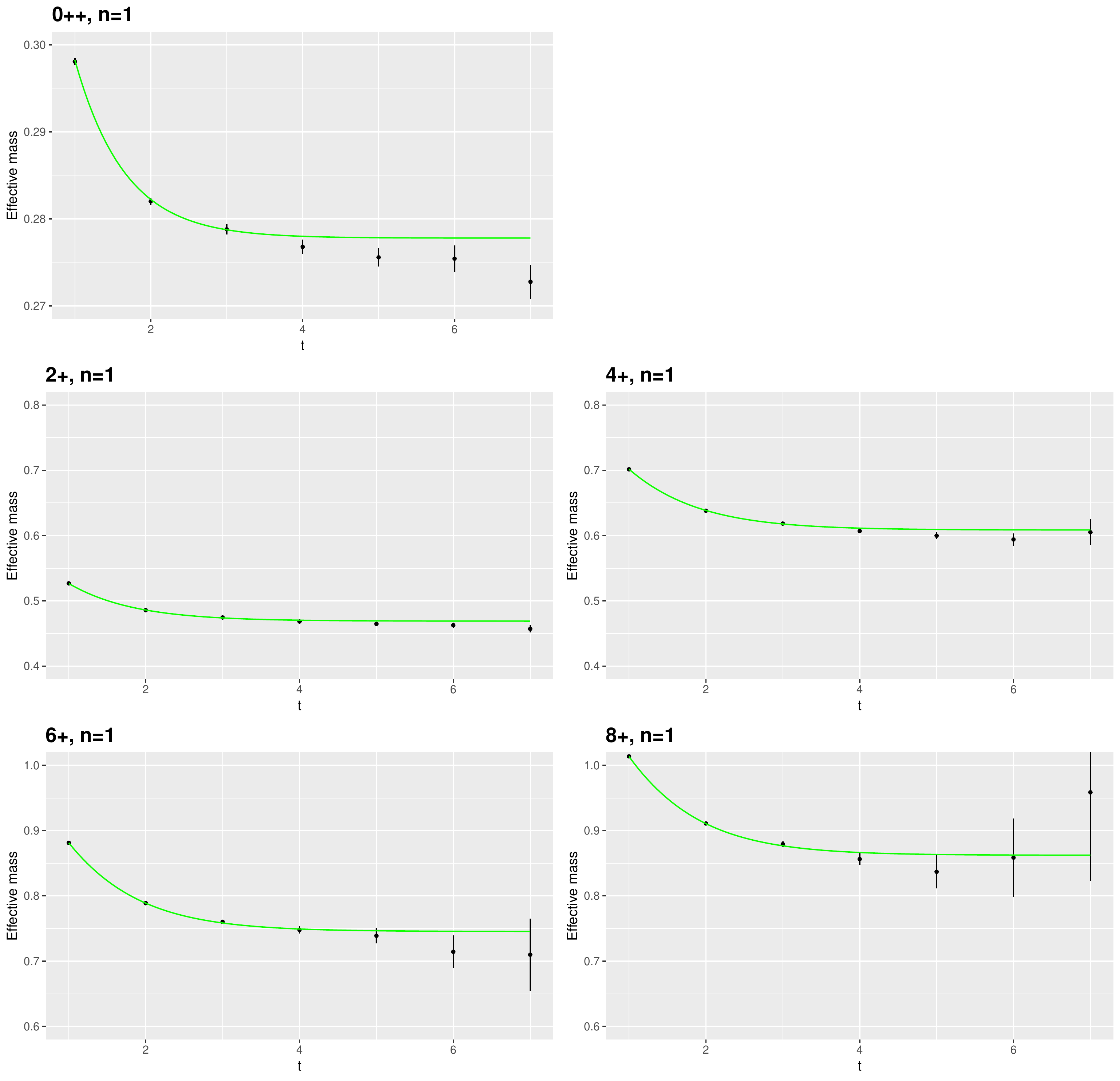}

\caption{Plot of effective mass $E_{eff}(t)$ versus $t$ (with fit, green curve)
for leading Regge trajectory states, $0^{++}$, $J^{C}=2^{+},4^{+}, 6^{+}, 8^{+}$. Note vertical scales differ between plots.\label{fig:massfitsRegge}}
\end{center}
\end{figure}

The rest of the paper is organized as follows. In Section~\ref{sec:ASA} we present a short summary of the ASA predictions for the glueball spectrum. We also compare them to the previously existing lattice results. In Section~\ref{sec:Methods} we describe in detail our methods.  The results are presented in Section~\ref{sec:results}, and also in Tables~\ref{tab:BadOverlaps}-\ref{tab:OldMassesA34-} and Figures~\ref{fig:Basic}-\ref{fig:Glueball-spectrum-byN-oneN5} which can be found at the end of the paper. We compare the measured glueball spectra with the ASA predictions and
find good agreement.
Future directions are discussed in the concluding Section~\ref{sec:conc}.

\section{ASA spectrum and earlier lattice data}
\label{sec:ASA}
Let us review the ASA predictions for the spectrum of $D=3$ glueballs. It is instructive to start with a brief reminder of what the spectrum of closed critical bosonic strings looks like (see, e.g., \cite{Polchinski:1998rq}).
The Hilbert space $\mathcal{H}_{c}$ of closed critical strings is a direct sum of subspaces labeled by a non-negative integer $N$, 
\begin{align}
\mathcal{H}_{c} & =\sum_{N}\mathcal{H}_{N}\;.\label{eq:H-sum}
\end{align}
For critical strings the masses of states at the same level $N$ are all equal to each other. This mass degeneracy follows from exact integrability of the worldsheet theory. Furthermore, each subspace $\mathcal{H}_{N}$
is a tensor product of spaces describing left- and right-moving worldsheet perturbations,
\be
\label{tensor_square}
\mathcal{H}_{N}=\mathcal{H}_{L}\left(N\right)\otimes\mathcal{H}_{R}\left(N\right)\;.
\ee 
The level number $N$ counts the total left-moving momentum along the string. The right-moving momentum necessarily takes the same value as a consequence of the worldsheet reparametrization invariance. This is known as a level matching condition.
 
 In order to define charge conjugation one needs $ \mathcal{H}_{L}\left(N\right)=\mathcal{H}_{R}\left(N\right)$, then
\be
\label{Caction}
C(\psi_L\otimes\psi_R)=\psi_R\otimes\psi_L\;.
\ee
A geometric meaning of (\ref{Caction}) is that charge conjugation acts by flipping the string orientation. 

Note that the tensor square structure (\ref{tensor_square}) follows from the absence of massive perturbations on the critical string worldsheet.
If present, these would allow for zero momentum excitations, modifying the tensor square structure.

Finally, for critical strings one finds that the spectrum of closed strings is related to the spectrum of open strings via
\begin{align}
\mathcal{H}_{L}\left(N\right) & =\mathcal{H}_{R}\left(N\right)=\mathcal{H}_{o}\left(N\right)\label{eq:H-LR-to-Hopen}\;,
\end{align}
where 
$\mathcal{H}_{o}\left(N\right)$ is the Hilbert space of open string states at level $N$.

According to the ASA confining strings at $D=3$ do not carry any massive excitations. Unlike in the critical case, the worldsheet theory is no longer  integrable \cite{Dubovsky:2014fma,Chen:2018keo}, so that the exact mass degeneracies at each level do not hold. However, as a consequence of approximate integrability, one still expects to find approximate level degeneracies. This leads to the first (weak) ASA prediction---glueballs are expected to come in 
approximately degenerate groups (levels), where each group can be decomposed as a tensor square (\ref{tensor_square}). As discussed in  \cite{Dubovsky:2016cog}, this ``weak" prediction is already quite restrictive. In particular, it implies that multiplicities of each level are equal to exact squares. Also, odd spin glueballs at each level appear in pairs of opposite charge parity and equal spin.

A stronger set of predictions follow from an assumption that the relation (\ref{eq:H-LR-to-Hopen}) continues to hold for confining strings. The structure 
of the open string Hilbert space ${\mathcal H}_o$ can then be described using the following semiclassical considerations. One starts with decomposing 
 ${\mathcal H}_{o}$ as a sum of subspaces ${\mathcal H}_j$ of a definite angular momentum $j$,
\be
\label{jsum}
\mathcal{H}_o  \equiv\sum_{N}\mathcal{H}_o\left(N\right)=\sum_{j\in\mathbb{Z}}\mathcal{H}_{j}\;.
\ee
Note that the sum in (\ref{jsum}) runs over all integer values of $j$. Open string states of spin $J>0$ are doublets of states with $j=\pm J$.
At large $J$ the minimum energy state in the corresponding $\mathcal{H}_{j}$ is described by a classical rotating rod solution. These states give rise to the leading Regge trajectory and satisfy $N=J$. Low lying excitations in $\mathcal{H}_{j}$ can be constructed by quantizing small perturbations around the rotating rod. This results in a prediction for  quantum numbers and multiplicities of states close to the leading Regge trajectory. 

The strongest set of the ASA predictions follow from an additional assumption that this perturbative analysis applies at {\it all} $j$ (including $j=0$ !) and
for arbitrarily high excited states in each $\mathcal{H}_{j}$. We will see that this somewhat {\it ad hoc}  assumption  works surprisingly well. Its possible justification may be that it is often the case that semiclassics produces exact results in integrable models. Confining strings are not integrable, however, according to the ASA they can be described as a deformation of an integrable theory. Discrete data, such as  state's multiplicities and quantum numbers, cannot change under a continuous deformation.

Adopting this assumption allows one to fully predict the multiplicities of states in each ${\mathcal H}_j$. Namely, the multiplicity of states
at the level $N$ in ${\mathcal H}_j$ is given by the $N$-th coefficient in the Taylor expansion of the following generating function
\be
P_{J}\left(x\right) = x^{J}\left(1-x\right)\mathcal{P}\left(x\right)\;,
\ee
where 
\[
\mathcal{P}(x)=\prod_{n=1}^{\infty}\frac{1}{1-x^{n}}
\]
 is the Euler generating function for integer
partitions.

The open string Hilbert space does not have direct physical meaning for  pure gluodynamics\footnote{It is natural to expect though, that it will correctly describe the meson spectrum if one adds to the theory a light fundamental quark flavor.}. To obtain  glueball quantum numbers one makes use of the tensor square decomposition (\ref{tensor_square}). The resulting predictions of the glueball quantum numbers for the first five levels ({\it i.e.}, for $N=0,\dots, 4$) are presented in 
Table~\ref{tab:ASA quantum numbers}. We see that the ``strong" ASA completely fixes quantum numbers of the first 39 glueball states (up to $N=3$). 

The ASA also fixes the quantum numbers of 64 glueball states at the $N=4$ level, up to a spatial parity assignment for two (out of ten) $J=0$ glueballs.
The latter ambiguity arises because the arguments above do not fix parities of $j=0$ open string states. This leaves parities of some $J=0$ glueballs undetermined, if there is more than one $j=0$ state at the corresponding open string level. $N=4$ is the first level when this happens. Independently of how this ambiguity is resolved, one of these states will be an exotic $CP=-1$ state, {\it i.e.} either $0^{+-}$ or $0^{-+}$. This makes the $N=4$ level especially interesting---this is the first place where such exotic state appears in the ASA spectrum.

To conclude this section and to set the stage for our further study let us review the status of the ASA spectrum as compared to the lattice data prior to the current work. A high precision mass determination for a large number of glueball states (and for up to $N_c=16$) has been recently performed in  \cite{Athenodorou:2016ebg}. However, $SO(2)$ spin values were not  measured in this simulation. Still, some limited information about glueball spins has been available. First, a number of low lying ${\mathbb Z}_4$ singlet states observed in    \cite{Athenodorou:2016ebg} do not have nearly degenerate partners of opposite parity. This implies that these are true $J=0$ states.
In addition, the lightest $J=4$ doublet has been identified in the earlier dedicated study presented in \cite{Meyer:2002mk,Meyer:2004gx}.
Furthermore, combining the results of \cite{Athenodorou:2016ebg}  with the earlier ones presented in \cite{Meyer:2003wx,Meyer:2004gx} it was argued in \cite{Athenodorou:2016ebg} that the preferred $J^C$ assignments for the four lightest odd spin glueballs are $1^-$ and $3^+$, $3^+$, $3^-$.

\begin{figure}[H]
\begin{center}
\includegraphics[width=1\textwidth]{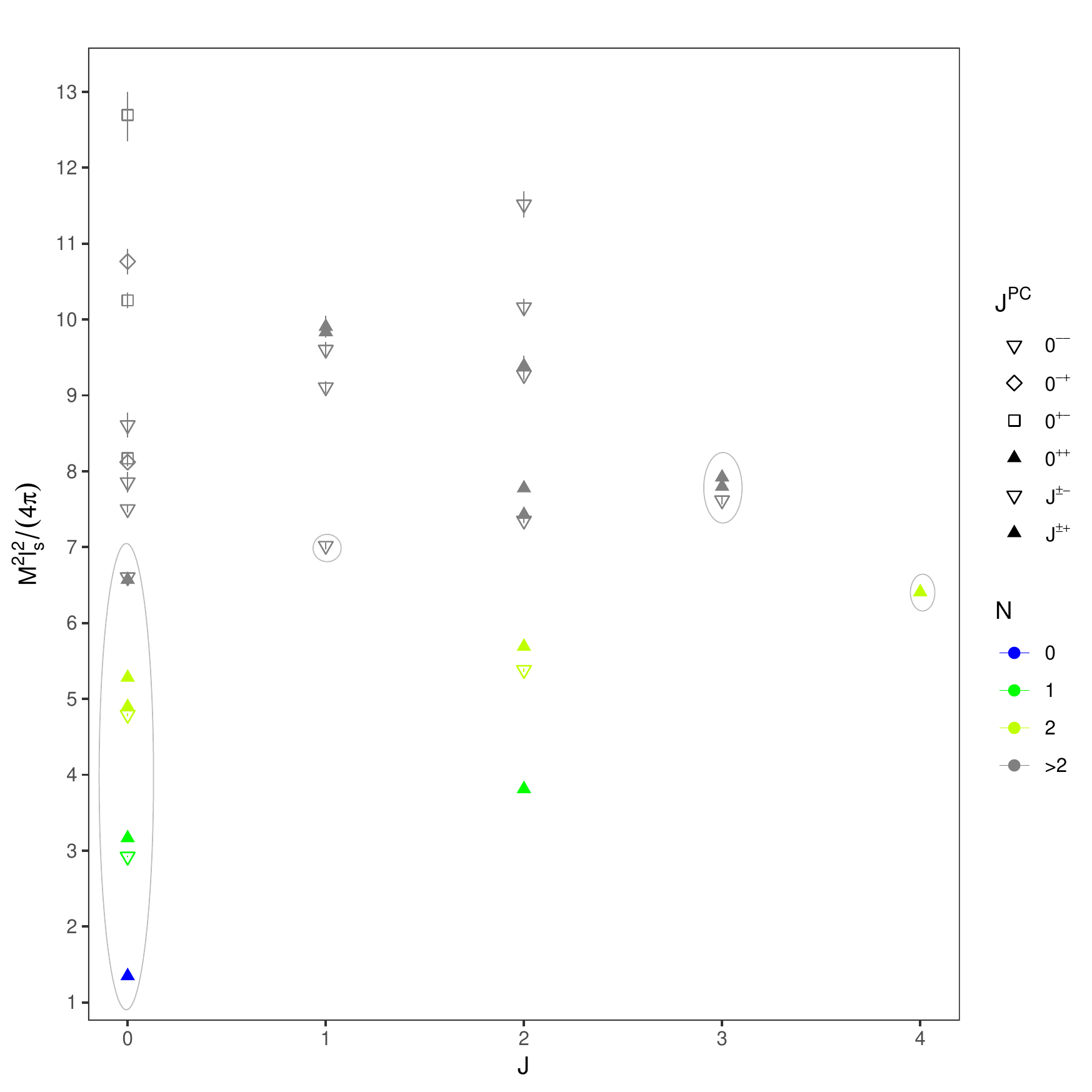}

\caption{Glueball spins determinations prior to this work; masses are $SU(\infty)$ limits or in a few cases  $SU(16)$ or $SU(12)$ masses from \cite{Athenodorou:2016ebg}. States circled had spin determined in \cite{Meyer:2002mk} or for $J=0$ states by lack of a candidate partner of opposite parity (see also \cite{Dubovsky:2016cog}). The string width $l_s$ is the inverse of the square root of the string tension $\sigma_{f}$.\label{fig:oldresults}}
\end{center}
\end{figure}

These results are summarized in Figure~\ref{fig:oldresults}. One immediately identifies in this figure well-pronounced $N=0,1,2$ levels perfectly matching the ASA predictions. 
This agreement is non-trivial, but should not be considered as a definitive confirmation of the ASA logic, given that the results presented in  Figure~\ref{fig:oldresults} were available prior to \cite{Dubovsky:2016cog} and served as a guide for the analysis presented in \cite{Dubovsky:2016cog}. Hence, the crucial test of the ASA should come from spin determinations for heavier glueballs with masses in the $N=3,4$ regions.
The goal of the present paper is to perform such a test. In particular, if the preliminary spin assignments for the four lightest odd $J$ doublets in  Figure~\ref{fig:oldresults} were correct, they would sharply contradict  even the weak ASA. Addressing this tension is one of the primary motivations for the current study.

\section{Methods}
\label{sec:Methods}
\subsection{Lattice simulations}

We generate configurations of $SU(4)$ matrices representing
lattice link variables, corresponding to exponentiated gauge fields,
in accordance with the standard Euclidean path integral
\begin{align}
Z & =\int\mathcal{D}U\exp\left(-\beta S\left[U\right]\right)=\int\mathcal{D}U\exp\left(-\frac{2N_{c}}{ag^{2}}S\left[U\right]\right),\label{eq:PI}
\end{align}
where $\mathcal{D}U$ is the Haar measure over the lattice link matrices,
$a$ the lattice spacing and $S$ is the Wilson plaquette action

\begin{align}
S & =\sum_{p}\left(1-\frac{1}{N_c}\textrm{Re}\textrm{Tr}U_{p}\right).\label{eq:Action}
\end{align}
Here $U_{p}$ is ordered product of link matrices around a plaquette
at location $p$.

We use a $80\times80\times96$ lattice  with $\beta=98.25$; 18 million
lattice configurations were generated using a standard Cabibbo-Marinari
heat bath plus over-relaxation algortihm, with operators being evaluated
every 100 configurations.

\subsection{Spin ambiguity and parity degeneracy on a cubic lattice}

As is well known, on a cubic $D=3$ lattice spatial Euclidean $SO(2)$
symmetry is broken to its ${\mathbb Z}_{4}$ subgroup\cite{Teper:PhysRevD.59.014512,Meyer:2004gx}.
Using only rotations of loop operators restricted to the lattice we
cannot distinguish the spins of eigenstates which have $J=0\:\textrm{mod}\:4$,
nor among those that have odds spins $J=1\:\textrm{mod}\:2$, nor among
those with $J=2\:\textrm{mod}\:4$. As discussed in detail in Section \ref{subsec:Augmenting-rotations}
we get around this problem by implementing  up to 16 approximate rotations
of loop operators, so that we can distinguish states of specific spins
up to $J=8$.

In three spatial dimensions $P$ parity can be defined as a reflection w.r.t. the origin. This transformation commutes with spatial rotations, so that the extended orthogonal group is simply a direct product
$O(3)=SO(3)\times {\mathbb Z}_2$, where the ${\mathbb Z}_2$ factor corresponds to $P$. On the other hand, in two spatial dimensions a reflection w.r.t. the origin is equivalent to a conventional rotation.
$P$ parity has then to be defined as a reflection w.r.t. one of the axis, for example the $y$-axis. As a result, $P$ does not commute with continuous rotations and the extended orthogonal group is a semidirect
(rather than a direct) product $O(2)=SO(2)\rtimes {\mathbb Z}_2$.
In other words,
the
parity and angular momentum operators do not commute.
A parity transformation switches clockwise and counter-clockwise rotations,
so that $P\ket{j}=\ket{-j}$. 

Now if $\ket{j}$ is a mass eigenstate
 we can form two opposite parity eigenstates $P\ket{j,\pm}=\ket{j,\pm}$ of the form
\begin{align*}
\ket{j,\pm} & =
\frac{1}{\sqrt{2}}\left(\ket{j}\pm\ket{-j}\right),
\end{align*}
{assuming}  both $\ket{j,\pm} $ are non-zero. Since $H$ and $P$
commute, these would then be degenerate mass eigenstates of opposite parity. For $j\neq 0$ the states $\ket{-j}$ and $\ket{j}$ are orthogonal so that 
\begin{align*}
\braket{\ensuremath{j,\pm|j,\pm}} & =1\;.
\end{align*}
Hence, $\ket{j,\pm} $  are indeed both non-zero at $j\neq 0$.

This continuum argument gets weaker on the lattice. It still applies for $J=odd$ states because $\mathbb{Z}_4$ rotations distinguish 
$j=1$  from $j=-1$ states, so that there is an exact parity doubling for odd spin states.
However, $2=-2 \,\textrm{mod}\,4$, so the argument does not apply to even spin states, and the parity doubling is broken by the lattice for these states. Of course, it does get restored in 
the continuum limit  $\beta\rightarrow\infty$ and $a\rightarrow0$.

This discusson implies that the irreducible representations of the symmetry group of the cubic lattice 
are $A_{1}$ (which correspond to the continuum states with $J^{P}=0^{+},4^{+},\ldots$), $A_{2}$ ($J^{P}=0^{-},4^{-},\ldots$),
$A_{3}$ ($J^{P}=2^{+},6^{+},\ldots$), $A_{4}$ ($J^{P}=2^{-},6^{-},\ldots$), and 
$E$ ($J^{P}=1^{\pm},3^{\pm},\ldots$).

\subsection{Approximately rotated operators\label{subsec:Augmenting-rotations}}

\begin{figure}[H]
\begin{center}
\subfloat[Set A\label{fig:Set-A}]{\includegraphics[width=0.9\textwidth]{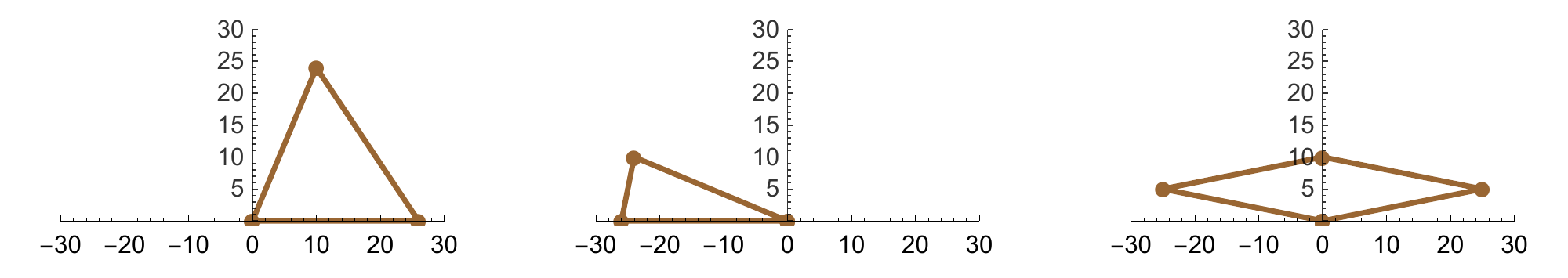}

}

\subfloat[Set B\label{fig:Set-B}]{\includegraphics[width=0.9\textwidth]{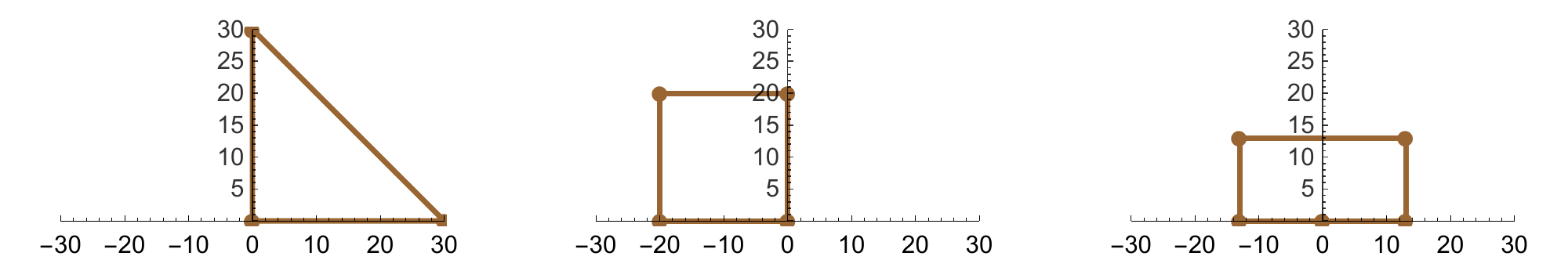}

}

\caption{Basic Wilson loop operators.}
\label{fig:Basic}
\end{center}
\end{figure}

\begin{figure}[H]
\begin{center}
\includegraphics{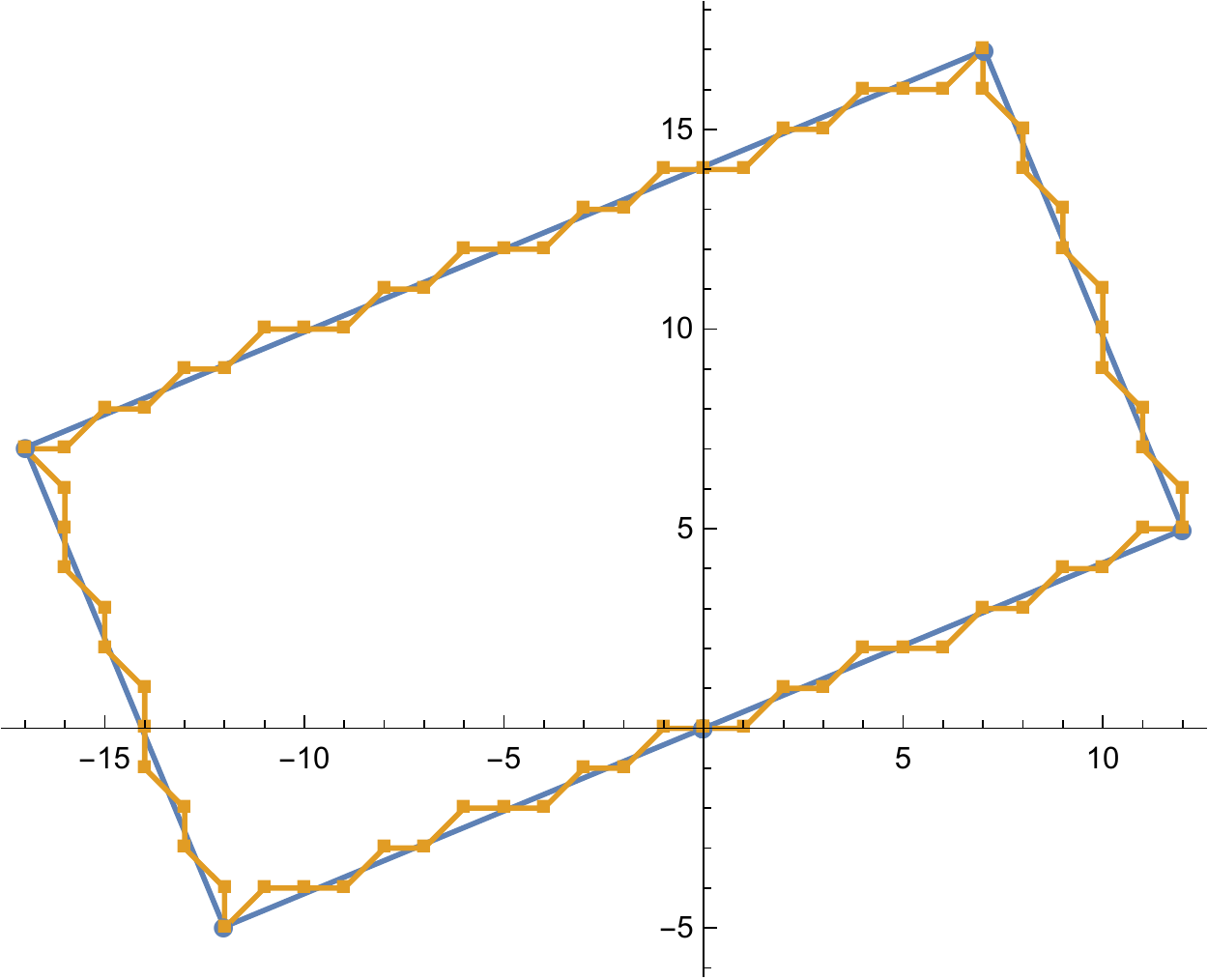}

\caption{Lattice approximation to third operator in Figure \ref{fig:Set-B}, rotated by $\pi/8$.\label{fig:Lattice-approximation} }
\end{center}
\end{figure}

\begin{figure}[H]
\begin{center}
\includegraphics[width=0.9\textwidth]{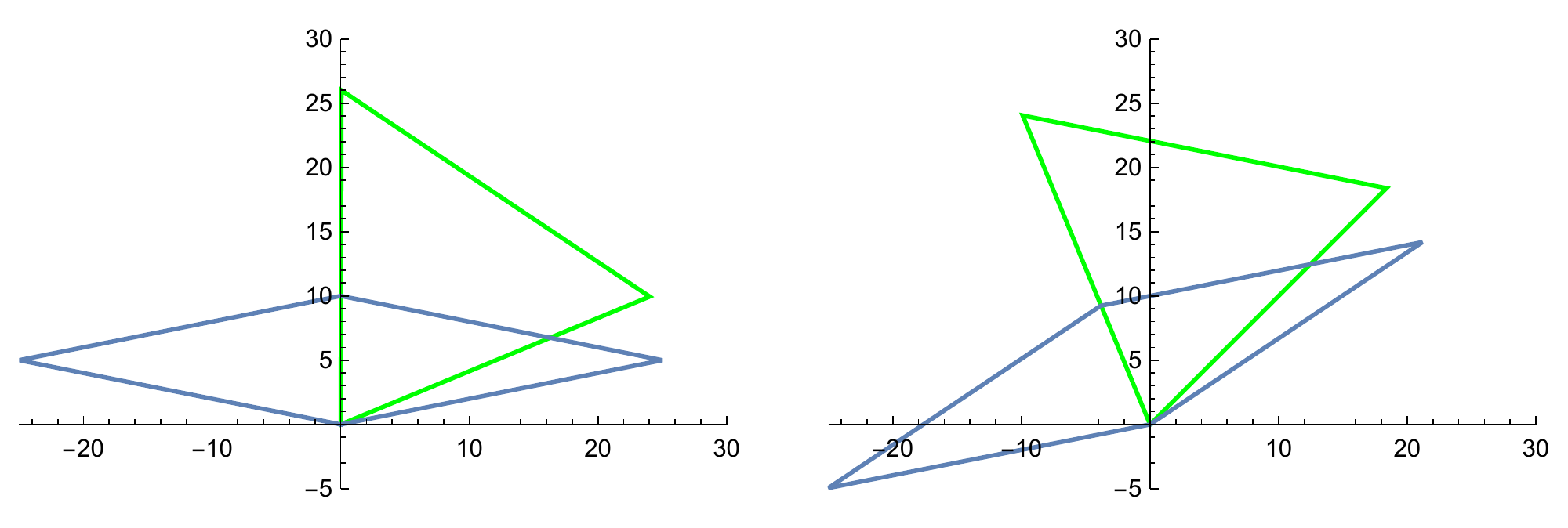}

\caption{Double operator based on connecting two of the operators in Figure
\ref{fig:Set-A}, unrotated and rotated by $\pi/8$. \label{fig:Double-operators} }
\end{center}
\end{figure}

In order to get around the breaking of the full rotational symmetry on a
cubic lattice, we make use of approximately rotated (AR) operators as
introduced in \cite{Meyer:2002mk}. Figure \ref{fig:Basic}
shows six basic operators that we use, while Figure \ref{fig:Lattice-approximation}
shows an example of an approximate rotation for one of these. An overall
physical scale of these operators is chosen to be similar to the
size (after blocking) of the operators used in \cite{Athenodorou:2016ebg}
which had large overlaps with light glueball mass eigenstates.

In order to construct AR operators we make use of additional gauge field matrices corresponding to diagonal links, as illustrated in 
 Figure \ref{fig:Lattice-approximation}. For instance, the diagonal link matrix in the $(1,1)$ direction is built by forming
\begin{align*}
 & \hphantom{=}U_{xy}\left(n_{x},n_{y},n_{t}\right)
 & =\frac{1}{2}\left(U_{x}\left(n_{x},n_{y},n_{t}\right)U_{y}\left(n_{x}+1,n_{y},n_{t}\right)+U_{y}\left(n_{x},n_{y},n_{t}\right)U_{x}\left(n_{x},n_{y}+1,n_{t}\right)\right)
\end{align*}
and by projecting the result back into  the space of $SU(N_c)$ matrices
exactly as it is done in the smeared link construction. Here $U_{x(y)}\left(n_{x},n_{y},n_{t}\right)$
is the gauge field matrix on the link, which starts  at lattice site $\left(n_{x},n_{y},n_{t}\right)$ and points 
in the spatial direction determined by the subscript. The definition of a diagonal link matrix in the $(1,$-1) direction is completely analogous.
Note that similarly to the smeared links, we don't use the diagonal links in the definition of the action, but only to construct a basis of AR operators.  Using these links we
calculated sixteen rotated
versions of each basic operator at angular increments of $\pi/8$. This allows construction of operators with (approximate) spin values $J$ from 0 up to 8. 

Clearly, the spin fidelity of rotated Wilson loops is better for
shapes that are larger in lattice units.
Keeping the physical size fixed, this requires smaller lattice spacing $a$.
 An important consequence
is that it is not feasible to construct small operators and then scale them 
up using blocking. We do use smeared operators.
However,  the inability
to implement  blocking  significantly restricts the quality of the overlap of the AR operator
basis with the light glueball mass eigenstates. 

Additionally, without blocking the calculation of the loop trace for
operators of this size in lattice units adds up to 100 multiplications
of $SU(N_c)$ matrices at each lattice point. In \cite{Athenodorou:2016ebg}
glueball masses were extrapolated to  $N_c=\infty$ values using masses obtained
for $N_c \in \left[2,16\right]$. However,  the authors of  \cite{Athenodorou:2016ebg} made use of blocking and also reduced the lattice size depending on $N_c$
to as small values as  $26^{2}\times 30$ for $N_c=16$. In the current work, the spin fidelity
of approximate lattice rotations would be unacceptable using a $26^{2}\times 30$
lattice size. 
In addition, each matrix multiplication requires $O\left(N_c^{3}\right)$
floating point operations. As a result, our mass determinations are limited to the 
 $SU(4)$ gauge group.

We further enlarge the AR operator  basis  by attaching pairs of basic operators at the origin 
as shown  in Figure \ref{fig:Double-operators}.
Since we already calculated  the product of link matrices around each
individual basic loop we can combine them  at the cost of just one more matrix
multiplication per lattice site for each double loop operator. For
any pair of distinct operators we can construct 16 distinct shapes
corresponding to different angles between the individual loops. For
each resulting double loop we then construct 16 rigid rotations of
that double loop. 

This procedure allows us to generate a large number of double loops at relatively small computational costs. Given
constraints on memory and processing time we restrict ourselves
to using double loop operators obtained by combining two of the loops in Figure
\ref{fig:Set-A} or two of the loops in Figure \ref{fig:Set-B}.
This gives us $2\times 96=192$ double loop operators before implementing rigid $\pi/8$ rotations.

Every operator is calculated after three and five smearings of lattice
links. Since we don't perform any blocking, after
a few smearings the values of the loop trace for a particular operator at different
smearing levels at a given lattice site become highly correlated.

Finally, in order to partially compensate for distortions due to the
approximate nature of  rotations, we renormalize each rotated
 operator by replacing it with 
\[
\bar\Phi=\frac{\Phi - \langle  \Phi \rangle}{\sqrt{\langle (\Phi - \langle  \Phi
    \rangle)^2  \rangle } }\;.
    \]

\subsection{Filtering of operators}
In
\cite{Meyer:2002mk} the rotational quality of operators was assessed
by examining the correlation matrix for rotations of a particular shape and checking that it is well approximated by a circulant matrix,  which it becomes in the continuum theory.
We take a different approach, dictated by a different use of AR operators in our analysis. Namely, we use them to construct operators with a definite value of $J$ via the angular Fourier transform.
These are used then to measure glueball masses in different $J$ subsectors.

Hence to test the rotational quality we look at correlations between different $J$ components
in the same ${\mathbb Z}_4$ representation. For every ${\mathbb Z}_4$ representation we reject 
operators for which
the absolute value of the overlap between any pair of different $J$ harmonics
 is greater than $0.02$. After variational analysis this
choice of threshold  is found to result in mass eigenstates
 that in most cases do not have significant overlaps between states of different $J^{PC}$.
 
 Nevertheless, we noticed several exceptions from this rule.  For example
Table \ref{tab:BadEigenOverlaps} shows overlaps between $0^{++}$
and $8^{++}$ states (for time separation 4 lattice units; see  Equation \ref{eq:overlapDefn}, Section \ref{subsec:Spectrum-using-conventional} for a precise definition of what we mean by overlap at some time separation), which are still statistically significant.
At some point in our analysis this effect gave
rise to a large systematic underestimation of $8^{++}$ masses. To avoid
this in the variational estimates for the mass eigenstates for each $J^{PC}$ we project out 
 components with smaller $J$ values in the
same $\mathbb{Z}_4$ representation and with the same $PC$ values. For example in the
case of $8^{++}$ we project out the overlap with $0^{++}$ and $4^{++}$.

 As an illustration of the rejection procedure,
 Table \ref{tab:BadOverlaps} shows the overlaps between the odd $J^{++}$
components of the second operator in Figure \ref{fig:Set-A}. Since
at least one of these is greater than $0.02$,
this operator is excluded from the variational analyses
 in the $1^{++}$, $3^{++}$, $5^{++}$ and $7^{++}$
representations. On the hand, the overlaps shown in Table \ref{tab:GoodOverlaps} for another operator
are deemed satisfactory and the corresponding operator is included the $1^{++}$, $3^{++}$, $5^{++}$, $7^{++}$  variational
analyses.

.

\subsection{Mass determination}
After constructing a set of $M$ operators $\phi_{i}$ with zero momentum
and the same definite $J^{PC}$, which are obtained via angular Fourier transforms, charge conjugation, reflections and spatial averaging
 of  particular loop shapes, we define
the cross-correlation matrix
\begin{align*}
C_{ij}\left(t\right)  =\langle\phi_{i}^{\dagger}(t)\phi_{j}(0)\rangle
  =\sum_{n}\langle vac|\phi^{\dagger}|n\rangle\langle n|\phi|vac\rangle e^{-E_{n}t}\;,
\end{align*}
where the sum is over mass eigenstates with the same $J^{PC}$, 
$t$ is time separation and $E_{n}$s are  eigenstate energies.
Given  lattice measurements for $C\left(t\right)$ for
different time lags $t$ we can then  recover the mass eigenstates
by solving the generalized eigenvalue problem \cite{Luscher:1990ck}
\begin{align}
\sum_jC_{ij}(t_{2})v^j_{n} & =\lambda_{n}\left(t_{1},t_{2}\right)\sum_jC_{ij}\left(t_{1}\right)v^j_{n},\label{eq:Variational}
\end{align}
 for $n=1,\dots M$. If we write 
 \[
 \psi_{n}=\sum_i v^{i\dagger}_{n}{\mathbf{\phi}_i}
 \]
and denote 
\[
c_{n}\left(t\right)=\langle\psi_{n}^{\dagger}(t)\psi_{n}(0)\rangle
\]
we then define an effective mass
\begin{align*}
E_{eff,n}(t) & =-\ln\frac{c_{n}(t)}{c_{n}(t-1)}.
\end{align*}
Of course for any $J^{PC}$ there is an infinite tower of mass eigenstates,
while we have only a finite set of operators.  As a consequence, each generalized
eigenstate $\psi_{n}$  obtained
from  (\ref{eq:Variational})  has non-vanishing overlaps with more than
one true mass eigenstate. However, as the time lag $t$ increases, the higher mass
components of a particular $\psi_{n}$ get exponentially suppressed. As a result at large $t$
the effective mass  $E_{eff,n}(t)$ approaches the mass of the lightest component
of $\psi_{n}$  (of course the $\psi_n$ are obtained by variational analysis at some short time separation and may include small components of lighter states, but these will  become material in correlators at larger time separations than we effectively fit in this paper).

For each $\psi_{n}$ we fit the normalized correlators ${c_{n}\left(t\right)}/{c_{n}\left(0\right)}$ with a sum of two exponentials 
\[
f\left(t\right)=w\exp\left(-m_{1}t\right)+\left(1-w\right)\exp\left(-m_{2}t\right)
\]
and take the minimum of two masses $\min\left(m_{1},m_{2}\right)$ as an estimate for the mass of the corresponding eigenstate.
 The fit is performed by minimizing
 the inner product 
 \[
 I=
 \sum_{j,k}\left(f\left(t_{j}\right)-\frac{c_{n}\left(t_{j}\right)}{c_{n}\left(0\right)}\right)\left(\Omega^{-1}\right)_{jk}\left(f\left(t_{k}\right)-\frac{c_{n}\left(t_{k}\right)}{c_{n}\left(0\right)}\right)\;,
 \]
where $\Omega$ is the jackknife covariance matrix of the normalized
correlators.

As discussed later,  a comparison of masses obtained with the AR operators with those obtained with the conventional operators
shows that the conventional masses are systematically lower.
This indicates that the AR operators have larger residual overlaps with heavy eigenstates. This is not surprising, given that 
no blocking is implemented for AR operators.
 In the future this problem can 
be addressed, for example, by fitting the normalized correlators by a sum of three or more exponentials.
Unfortunately, with the currently available amount of data,
 this gives rise to very large statistical errors
which make it impossible to extract the masses to a useful precision.

\section{Glueball spectrum}
\label{sec:results}
\subsection{Spectrum with AR  operators vs $N=3$ ASA\label{subsec:Spectrum-using-enhanced}}
Applying methods described in Section~\ref{sec:Methods} we perform mass and spin determinations for a large number of glueball states.
The results of these measurements are summarized in a number of Tables and Figures at the end of the paper.

\begin{figure}[H]
\begin{center}
\includegraphics[width=1\textwidth]{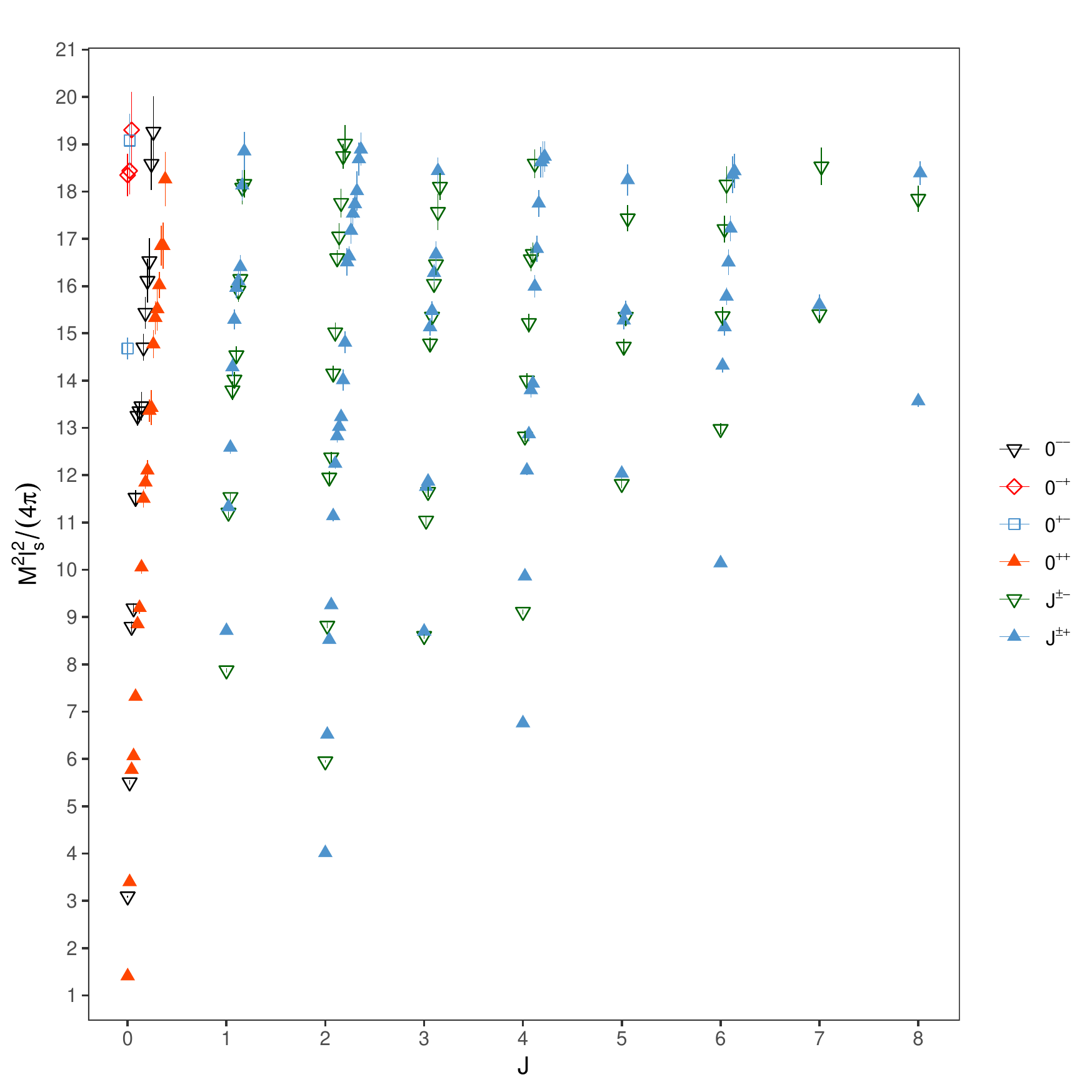}

\caption{The Chew-Frautschi plot.\label{fig:Glueball-spectrum}}
\end{center}
\end{figure}

In Tables \ref{tab:MassesA1A2},
\ref{tab:MassesEbyP} and \ref{tab:MassesA34byP} we present the mass values in lattice units for 15 lightest states in each channel where we perform the analysis.
In Figure \ref{fig:Glueball-spectrum} we show the same glueball spectrum
in a Chew-Frautschi plot (with $M^2$ and $J$ axes swapped as compared to how it is usually presented), 
where the squared mass has been expressed relative to the string tension $\sigma_{f}=l_s^{-2}$, where $l_s$ is the string width.   $\sigma_{f}$ is determined in a conventional manner from the  ground state temporal correlator and corresponding energy for fundamental flux tubes winding once round the lattice in the
spatial direction \cite{Athenodorou:2016ebg},
\cite{Teper:PhysRevD.59.014512}) for the value of $\beta=98.25$ used
here and the square root is found to be 0.06603(8). 
For $J>0$ this Figure shows for each $J^{C}$ only the masses averaged over two $P$ values\footnote{As expected pairs of masses for $P=\pm$ for given $J^{C}, J>0$ typically agree within statistical errors in Tables \ref{tab:MassesA1A2}, \ref{tab:MassesEbyP} and \ref{tab:MassesA34byP}. There are exceptions, for example the ground states for $8^{++}$ and  $8^{-+}$, but as we discuss in  Section \ref{subsec:Spectrum-using-conventional} systematic errors are anyway greater than statistical errors for the AR operators at the corresponding high masses.}. 
Also, in this Figure we cut the spectrum at high masses at $m=1$ in lattice units.
The errors  indicate 1$\sigma$ statistical
uncertainties only.
 As we discuss below, in most cases systematic errors  are quite a bit larger than the statistical ones (typically, systematic errors result in masses presented in Figure \ref{fig:Glueball-spectrum} being heavier than the actual physical masses). 
 
 For now, let us neglect possible systematic errors and  compare these results with the ASA predictions  (see Table \ref{tab:ASA quantum numbers}) for up to $N=3$ string level, taking the mass values (or rather, the ordering of masses) presented in Figure \ref{fig:Glueball-spectrum} at face value.
  Most conveniently this can be done  using Figure \ref{fig:Glueball-spectrum-byN-toNequals3}, where we colored the states according to the ASA level assignment up to $N=3$, and cut the heavier states by keeping only one heavier $N=4$ state in each $J^{PC}$ channel. 

  Remarkably, we observe a perfect match to the ASA predictions for the $N=3$ states, in addition to the previously observed match for $N=0,1,2$. In particular, the sharp conflict between earlier preliminary spin determinations for four low lying odd $J$ states and the weak ASA predictions got resolved---our spin values for these states are in a complete 
agreement with the ASA. We see a well-pronounced  group of $1^\pm$ and $3^\pm$ ground states separated by a large gap from heavier states in the respective channels. In Figure~\ref{fig:MassplotExample} we also presented the effective mass for each of these four states as a function of the time lag in lattice units together with the corresponding fits with two exponentials. 

This achieves the  minimal goal for the current study. We see that the ASA does provide a convincing description of glueball quantum numbers below the  $N=4$ level.
Now, given the large amount of spin determinations for heavier states, it is natural to check the ASA predictions for the  $N=4$ states as well.
Before doing this, let us take a closer look at the systematic errors in our data.

\subsection{Spectrum with conventional operators and systematics\label{subsec:Spectrum-using-conventional}}

As a way to cross-check our results and to assess the involved systematics it is natural to compare the AR spectrum with the one based on a conventional set of operators. To achieve this we measure
 glueball masses from the same set of gauge field configurations
using one of the operator bases used in \cite{Athenodorou:2016ebg}.
Of course, spin identifications for the latter states are subject to the usual limitations
of square lattices.

 The resulting masses  are presented in
Tables \ref{tab:OldMassesA1+} to \ref{tab:OldMassesA34-}, where we show first 20 states for each set of ${\mathbb Z}_4$ and $P,C$ quantum numbers. In addition
we show the  normalized overlaps between each corresponding generalized eigenvector and
the generalized eigenvectors in different $J$ channels obtained with the AR operator set. Here by 
a normalized overlap between operators $\psi$ and $\phi$ at time separation  $t$
 we understand
\begin{equation}
O(\psi,\phi,t)=\frac{\langle\psi^{\dagger}(t)\phi(0)\rangle}
{\sqrt{\langle\psi^{\dagger}(t)\psi(0)\rangle\langle\phi^{\dagger}(t)\phi(0)\rangle}}\;.\label{eq:overlapDefn}
\end{equation}
As a consequence of this definition at late time one finds
\[
O(\psi,\phi,t\to \infty)=1
\]
if the lowest energy eigenstates appearing in the decompositions of $\psi|0\rangle$ and $\psi|0\rangle$ in the energy eigenbasis are the same and non-degenerate.
Overlaps shown in Tables \ref{tab:OldMassesA1+} to \ref{tab:OldMassesA34-} are defined by 
\[
O_{20}(\psi,t)=\sqrt{\sum_{i=1}^{20}O(\psi,\phi_i,t)^2}\;.
\]
where the sum goes over 20 lightest states in each corresponding $J$ channel\footnote{Note that this number is larger than 15 states shown previously in Tables \ref{tab:MassesA1A2},
\ref{tab:MassesEbyP} and \ref{tab:MassesA34byP}, where we chose to cut states earlier in order to avoid an excessive cluttering.}.
For example, in Table \ref{tab:OldMassesA1+} we show overlaps
between each $A_{1}$, $C=+$, conventional state with
the $0^{++}$, $4^{++}$ and $8^{++}$ AR states and finally the total overlap
with all the latter. In all cases we show overlaps at two different times, $t=0$ and $t=4$ in lattice units.

We see that for most states the total overlap does not add up to unity. 
Most likely the main cause of this is that  both AR and conventional operators  have a substantial admixture of heavier states, which still contribute at $t=4$. In agreement with this interpretation total overlaps get considerably closer to unity  at $t=4$ as compared to $t=0$. Note that the masses obtained with conventional operators are in a good agreement with the results of \cite{Athenodorou:2016ebg}, while the masses for the same states obtained with the AR operators are quite a bit higher. This indicates that the AR operators have larger admixture of heavy states, as expected, because no blocking was applied for them.

For all states with mass less than 0.80 in lattice units we make a tentative
assignment of spin based on overlaps at separation $t=4$. In the Tables, the
corresponding ``dominant" overlap is denoted in colored bold or italic.
In many cases this assignment is unambiguous, meaning that the corresponding state indeed contains a clear dominant component with some value of $J$. However, for a number of states this is not the case, and generalized eigenstates have comparable components with different values of $J$ when decomposed in the AR basis.  This happens when there is a group of states with different spins and close values of the mass. In these cases the generalized eigenstates obtained with conventional operators are actually admixtures of several states with different $J$. Even in these cases usually the overall count for the number of states is unambiguous, even though an assignment of a definite $J$ value to a specific conventional mass eigenstate does not have well-defined meaning.

For example, a situation like this arises for the two lightest $E^+$ states. As follows from Table \ref{tab:OldMassesE+} one finds here two nearly degenerate states with a significant admixture of both $J=1$ and $J=3$ (even though in both cases one of the components is considerably larger than another). Still, the conclusion that there is one true eigenstate with $J=1$ and another with $J=3$ at this mass is robust, in agreement with the ASA predictions and the analysis in the AR basis. On the other hand, as can be seen from Table \ref{tab:OldMassesE-}, the mass gap between the two lightest $E^-$ states is larger and no ambiguity of this kind arises there.

At high spins the conventional basis starts missing states. Neither a clean $J=8$ candidate  in Tables \ref{tab:OldMassesA1+}, \ref{tab:OldMassesA2+}, \ref{tab:OldMassesA1-}, \ref{tab:OldMassesA2-}
nor a clean $J=7$ candidate in Tables \ref{tab:OldMassesE+}, \ref{tab:OldMassesE-} are found. This problem is present even at $J=6$---what we marked as the lightest $J=6^+$ state in Table~\ref{tab:OldMassesA34+} is one of two nearly degenerate states with a significant $J=6$ component (with another such state also being closeby), but for all of these states the $J=2$ component dominates.
This is not surprising given that conventional operators have only been rotated by multiples of $\pi/2$, so their overlaps with highly oscillating large $J$ states are suppressed.

\begin{figure}[H]
\centering
\includegraphics[width=0.7\textheight]{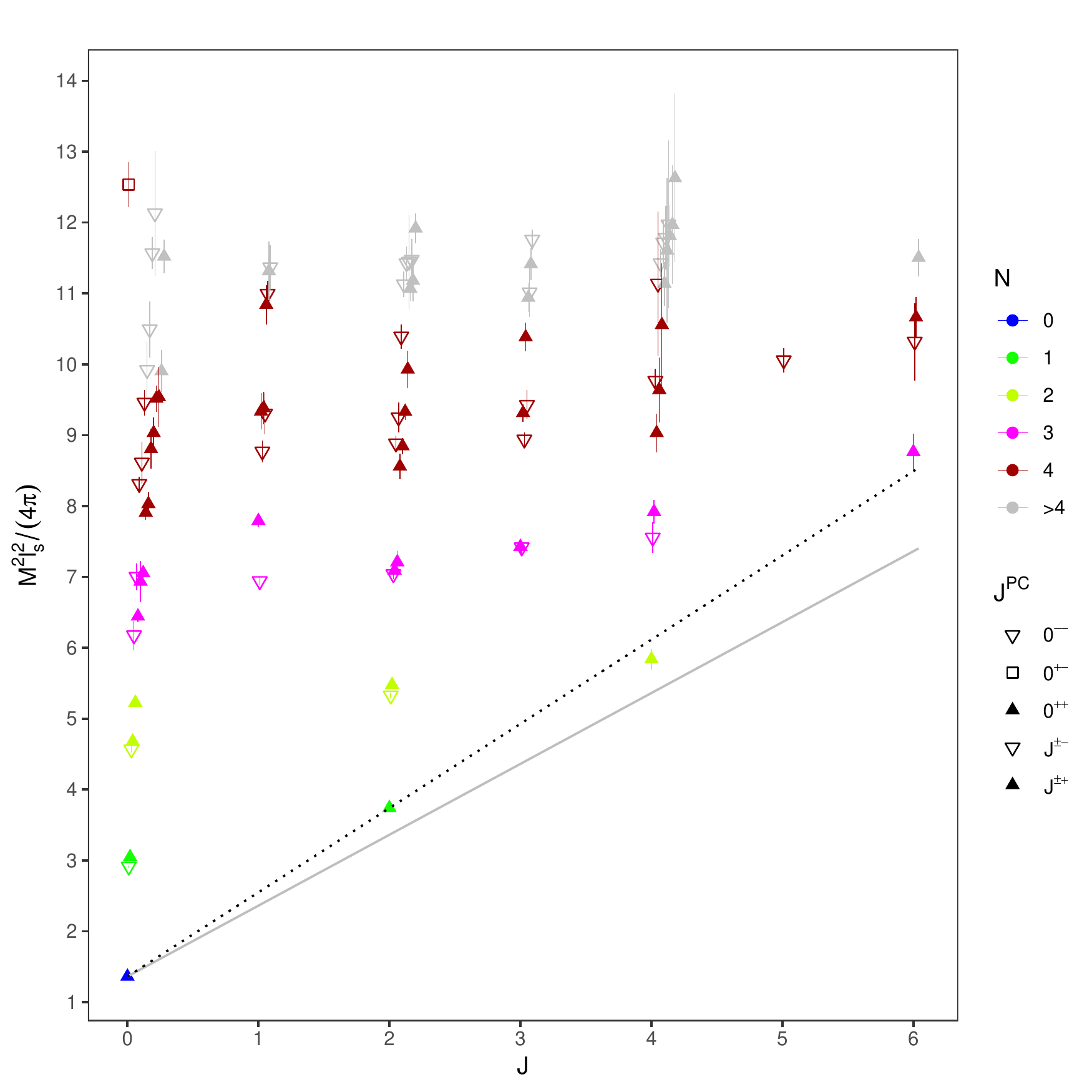}
\caption{The Chew-Frautschi plot with color coded level assignment; masses  are obtained using conventional
 operators and the $J$ assignments are based on overlaps with
AR eigenstates. The dotted line is the weighted least squares
fit to the leading Regge trajectory ($M^{2}l_{s}^{2}=4\pi\left(1.362(6)+1.19(1)J\right)$).
The light solid line
is the linear Regge trajectory with a unit slope line starting at $0^{++}$ ground state.\label{fig:Glueball-spectrum-old-ops}}
\end{figure}

 With all these caveats in mind, we have then counted
a number of states in each spin channel from the lowest to the highest mass and for
$N={\color{blue}0},{\color{green}1},{\color{lime}2},{\color{col3}3},{\color{col4}4}$
identified the corresponding ASA level by color coding superimposed
on bold font. States in italic have $N>4$ according to the ASA. Figure \ref{fig:Glueball-spectrum-old-ops} shows all these states in 
the Chew-Frautschi plot. We again find a well pronounced $N=3$ level with the same spin content as observed with the AR basis (barring the above subtlety with identification of the $J=6$ state).

As we already said, the AR masses are systematically heavier than the conventional ones.
It is natural to use this as an estimate for systematics in the AR mass determinations.
To illustrate this effect in Figure \ref{fig:Spectrum_with_systematics} we plotted
the AR masses with additional dashed  bars extending down to the corresponding 
point in Figure \ref{fig:Glueball-spectrum-old-ops}.  As expected, 
 systematic errors increase with mass. 
 By the time one reaches the $N=4$ level they become larger than the level separation.

Another characterization of systematic errors may be obtained by fitting the leading 
Regge trajectories for both sets of masses. These can be approximated by 
\[
M^{2}l_{s}^{2}=4\pi\left(1.40(4)+1.35(4)J\right)
\]
for the AR operators and by 
\begin{equation}
M^{2}l_{s}^{2}=4\pi\left(1.362(6)+1.19(1)J\right)\label{eq:Regge-old}
\end{equation}
for the conventional ones.
The greater slope in the AR case reflects the increase in systematic errors at larger masses.
Note that at large $J$ the trajectory should be well described by semiclassical folded rotating rod solution and the slope should approach $4\pi$.
Even with conventional masses the slope is larger than this value by a factor 1.19. Of course, one does not expect the exact agreement at lowest spins, but  it is apparent from Figure \ref{fig:Glueball-spectrum-old-ops}
that the shape of the observed trajectory does not go in the right direction as the spin increases from $J=4$ to $J=6$. The most likely explanation for this is the presence of a substantial systematic error in the measured mass of the $J=6$ state even with the conventional basis of operators. 

We show effective masses and sum of exponential fits for the leading Regge trajectory states obtained from the AR operators in Figure~\ref{fig:massfitsRegge}.

\subsection{Finite volume and discretisation errors}
\label{section_V&aerrors}

Our calculations have been performed on a single volume and at a single
value of the lattice spacing. So an important question is whether the
finite volume and lattice spacing corrections are small enough that they
do not affect our conclusions. Our choice of volume was guided by the
finite volume checks in earlier work
\cite{Athenodorou:2016ebg}
and our lattice spacing is smaller than the smallest lattice spacing
used in that work. However the calculations in
\cite{Athenodorou:2016ebg}
involved a smaller number of states
than in the present work and there was no attempt to identify
the real continuum spins of the states involved. In particular one
might expect that states of higher spin, which require higher angular
resolution for their identification, might have larger $O(a^2)$
discretisation corrections, and this is one reason that
we performed our calculations at such a small value of $a$.
In addition to this, states of higher spin should
be larger (as will be some of the more massive states of lower spin)
and this may lead to larger finite volume corrections. 
The only unambiguous way to deal with these issues is to
repeat our calculations for a range of values of $\beta\propto 1/a$,
and also for a range of spatial volumes. Unfortunately this ideal
approach would involve a much larger calculation which is beyond the
scope of the present paper. Instead what we shall do here is to return
to some of the calculations in 
\cite{Athenodorou:2016ebg}
and focus on a selection of states that have particular relevance to
our present work.

We focus on the $SU(4)$ calculations at the three largest values of
$\beta$ in
\cite{Athenodorou:2016ebg},
i.e. $\beta=63,74,86$. Since masses vary as $am\propto O(1/\beta)$ the
$O(a^2)$ lattice spacing corrections should decrease by a factor
$\sim 0.5$ when we go from $\beta=63$ to $\beta=86$, so any
large lattice corrections will lead to visible changes in 
ratios of masses as we vary $\beta$ over this range. The values
of the mass gap and string tension are listed in
Table~\ref{table_mM}. The larger spatial size, $l=70a$,
at $\beta=86$ satisfies $lM_{0^{++}}\simeq 21.8$ as do the lattice
sizes at $\beta=74$ and $\beta=63$. The lattice size in the
present paper, $l=80a$ at $\beta=98.25$, is roughly the same. In
Table~\ref{table_mM} we have also included calculations at $\beta=74$
on a smaller lattice size so as to investigate finite volume
corrections. All these calculations have been performed with the
same ``conventional" basis of operators used in the present
paper. These operators project onto states in the irreducible
representations of the square symmetry, so we do not know what
their continuum spins are, but we can use what we have learned
in Tables~\ref{tab:OldMassesA1+} to \ref{tab:OldMassesA34-} to determine which of these
states are interesting for our purposes.

To compare the mass of a state at different $\beta$ (and volumes)
we compare the ratio of this mass to that of the
lightest scalar glueball (the mass gap). (As is evident from
Table~\ref{table_mM} the mass gap, expressed in units of the
string tension, is independent of $\beta$ within the small
statistical errors.)

The first states we choose to focus upon are the lightest 6 in the
$A_4$ $C=+$ representation. The predictions of the ASA in
Table~\ref{tab:ASA quantum numbers} are that these
states include 4 $J^{PC}=2^{-+}$ states and 1 $J^{PC}=6^{-+}$
state from the $N=1,2,3$ levels, and one further state from the
$N=4$ level. The calculated overlaps in Table~\ref{tab:OldMassesA34+}
support this spin identification (with some possible mixing).
We observe from Table~\ref{table_mM}
that for all of these states any finite volume corrections 
cannot be much larger than the small statistical errors.
Any lattice spacing corrections are also invisible except possibly
for the first excited state, but even here any shift is far too
small to alter our conclusions about the ordering of the lightest
levels. 

The second set of states we consider are the lightest 2 in the $A2$
$C=+$ representation, which we know from Table~\ref{tab:OldMassesA2+}
to be states with $J^{PC}=4^{-+}$. Here we see no significant
finite volume corrections, and any finite lattice spacing corrections
are small and within about two standard deviations.

Finally we consider the lightest 3 states in each of the $E$ $C=+$
and $C=-$ representations. From
Table~\ref{tab:ASA quantum numbers} we expect the lightest
two states to be $J=1$ and $J=3$ and that they will be
close in mass since they both belong to the $N=3$ level. The
gap to the next state should be larger since it belongs to the
 $N=4$ level. This expectation is supported by the overlap
analysis in Tables~\ref{tab:OldMassesE+},~\ref{tab:OldMassesE-}.
We see from Table~\ref{table_mM} that these states suffer no
significant finite volume or lattice spacing corrections.

This comparison of a number of states selected from the $N=1,2,3,4$
levels suggests that our calculations and conclusions in this paper
are unaffected by significant finite volume and lattice spacing
corrections. This check includes a few states with $J=4, 6$ but
cannot, of course, reassure us about the states of higher spin
considered in this paper. For example, $J=8$ appears as too
high an excitation in the $A_1$ or $A_2$ spectra to allow an unambiguous
matching between spectra at various $\beta$ or volumes using
just our conventional operators. A larger, dedicated calculation
is clearly desirable.

\subsection{Lattice data vs $N=4$ ASA}
\begin{figure}[H]
\includegraphics[width=1.0\textwidth]{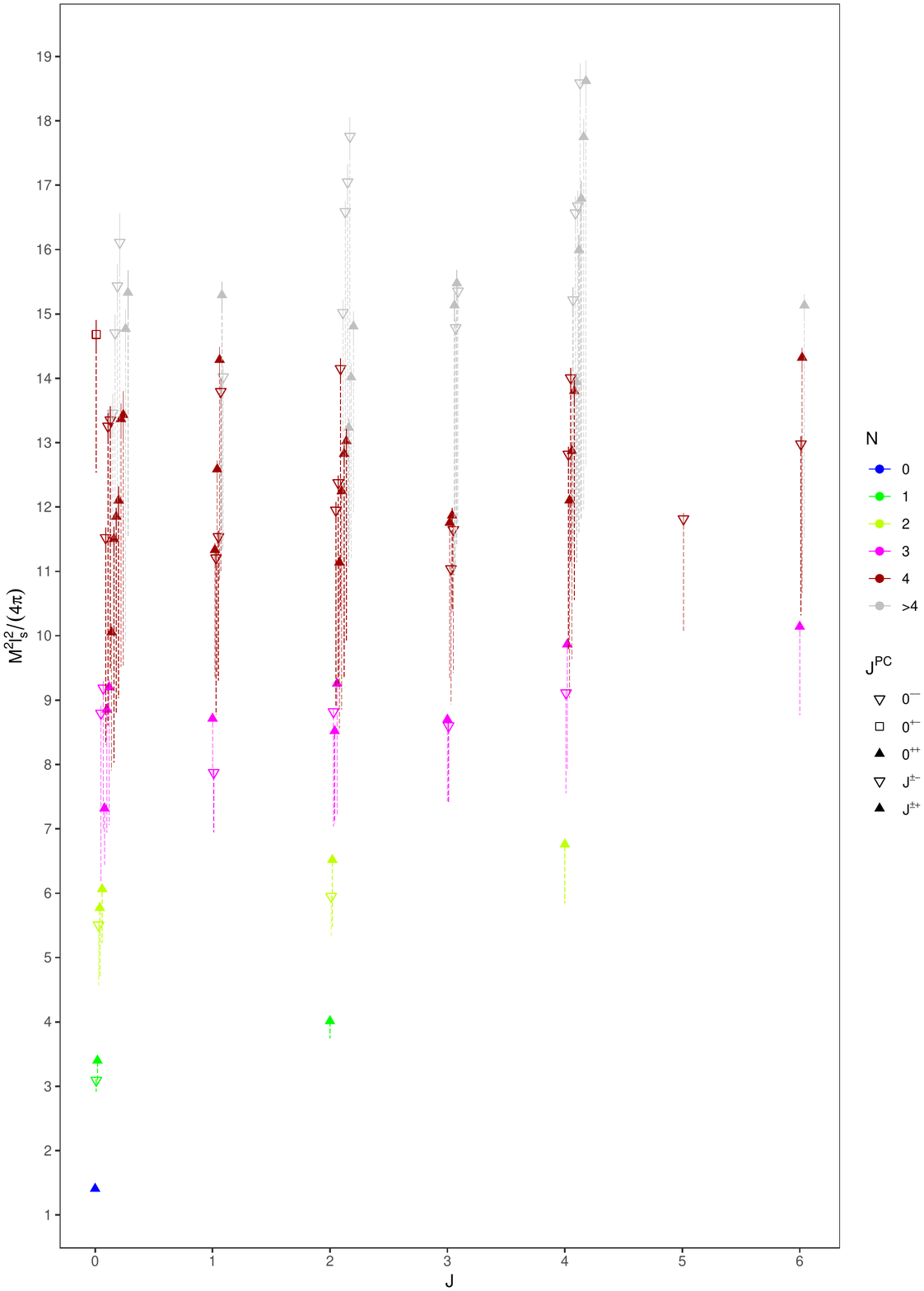}

\caption{Glueball spectrum with AR masses. Dotted  bars extend till the corresponding conventional mass value. \label{fig:Spectrum_with_systematics}}
\end{figure}

As should be clear from the results presented in Section~\ref{subsec:Spectrum-using-conventional} and summarized in Figure \ref{fig:Spectrum_with_systematics}, at the current stage the measurements with the AR operator basis are not to be considered  as high precision determinations of glueball masses. On the other hand, these measurements do provide a reliable determinations of glueball spins and of the relative order of glueball states, at least for $N<4$ levels. Figures  \ref{fig:Glueball-spectrum-old-ops} and  \ref{fig:Spectrum_with_systematics} confirm our earlier conclusion that the ASA provides the correct description of glueball quantum numbers at $N=0,\dots, 3$ levels.

\begin{figure}[H]
\begin{center}
\includegraphics[width=1\textwidth]{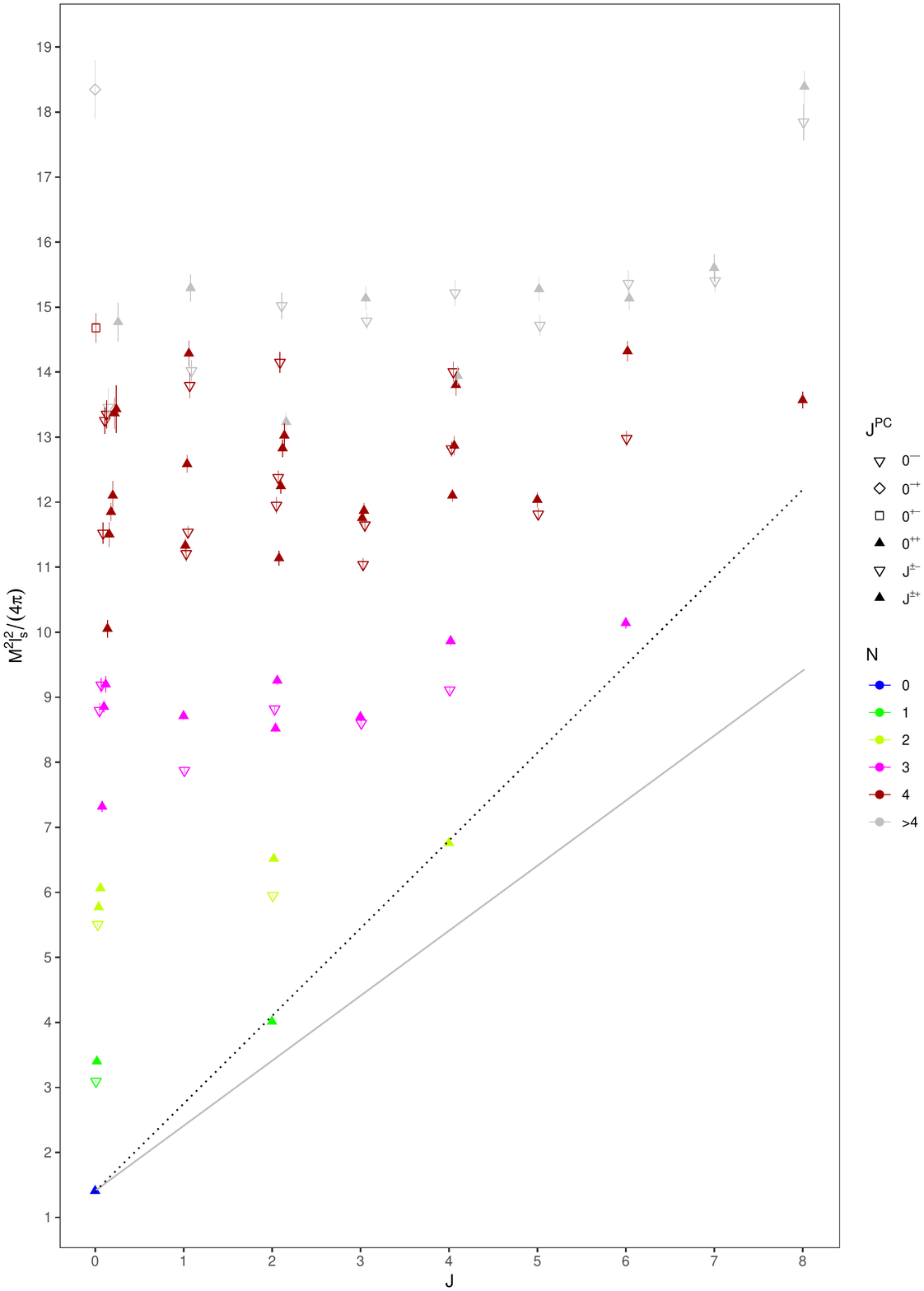}

\caption{The Chew-Frautschi plot  with color coded level  assignment up
  to $N=4$.  
\label{fig:Glueball-spectrum-byN-oneN5}}
\end{center}
\end{figure}

Let us now discuss what can be said about the $N=4$ states. Of course, if one literally took  the shifts in Figure \ref{fig:Spectrum_with_systematics}  as an  estimate of systematic errors, it would be very hard to distinguish between $N=4$ and $N=5$ states. However, these  shifts move all the states downwards. Even though the shifts may vary depending on a state or a symmetry channel, still it looks plausible that the AR spectrum 
correctly captures relative ordering of glueball states in the range of masses corresponding to the $N=4$ level and may be used there to assess the validity of the ASA. This is most conveniently done by using Figure~\ref{fig:Glueball-spectrum-byN-oneN5}, which is a direct analogue of Figure~\ref{fig:Glueball-spectrum-byN-toNequals3}, but moving one level higher.

We feel that Figure~\ref{fig:Glueball-spectrum-byN-oneN5} and Table~\ref{MassesByN} demonstrate an overall good agreement between the $N=4$ ASA predictions and lattice data. At some spin values  (such as $J=8,7,5,3$ and perhaps $J=6$) the agreement is in fact very convincing. At the remaining spin values one finds some ``contamination" of the $N=4$ level by putative $N=5$ states. It seems premature to conclude that these contaminations are indicative of a strong tension between the ASA and the lattice data before more precise mass determinations with the AR basis become available.

One should also keep in mind that the number of glueball states rapidly grows with $N$, while the level separation stays constant in $m^2$ units. For instance, at $N=5$ level the ASA predicts
\be
(0^{P_1}+0^{P_2}+1+1+2+3+5)^2=144
\ee 
states (their quantum numbers can be easily worked out using the $O(2)$ ``multiplication table" presented in Eqs.~(9)-(13) of \cite{Dubovsky:2016cog}). As a result, one expects the levels to become broader and to start overlap at large $N$, making the ASA quantum numbers predictions not very useful for many channels. Note that the ``clean" spin values in  Figure~\ref{fig:Glueball-spectrum-byN-oneN5} are also the ones with a smaller number of states, suggesting that we are starting to see the onset of this effect. This expected broadening of high levels does not contradict the approximate integrability of the worldsheet theory at high energies assumed in the ASA. Indeed, a typical large $N$ glueball correspronds to a long string state with a broad spectrum of excitations, so that the spectrum of heavy glueballs does not provide a direct probe of the high energy worldsheet dynamics.

The broadening does not make ASA completely useless at large $N$, because one can still follow its validity in exclusive less ``crowded" channels. A very interesting example of such a situation happens at $N=4$ level. This is the first time the ASA predicts an appearance of  (one) exotic spin 0 state with $CP=-1$. One indeed finds an isolated $0^{+-}$ state in Figure~\ref{fig:Glueball-spectrum-byN-oneN5}. Its AR mass is somewhat  heavier than other $N=4$ masses, but not badly so. With the current quality of data we find that  the very appearance of such isolated exotic state in the right ballpark of masses is encouraging.

Note that using the conventional operators in Figure~ \ref{fig:Glueball-spectrum-old-ops} one also finds an isolated $0^{+-}$ state but it appears to be separated further from the rest of $N=4$ states (as far as counting of
``contaminating" $N=5$ states goes). However, by inspecting Table~\ref{tab:OldMassesA1-} one finds that this state is a member of an extended group of $A_1^-$ states with relatively large and overlapping statistical error bars, so it well may be that the $0^{+-}$ state is misidentified in  Figure~ \ref{fig:Glueball-spectrum-old-ops} and a lower member of that group should be used instead.

\section{Conclusions and Future Directions}
\label{sec:conc}
To summarize, the main conclusion from the presented results is that the ASA provides an accurate description of the glueball spectrum in $D=3$ YM at least at the $N=0,\dots 3$ string levels, which correspond to the first 39 glueball states. It also provides, at the very least, a good organizing principle to describe the glueball spectrum as a whole. There is a number of directions to improve and extend these results both on a lattice and on a theory side.

On a lattice side it will be important to reach the stage when the AR basis can be used for high precision mass spectroscopy. This requires a significant reduction in the systematic overestimate of the glueballs masses which we observe with the present AR basis. This goal may be achieved with future simulations with a larger basis of AR operators and with higher statistics. In addition, for a cleaner test of the ASA, it is important to extend the analysis to a larger number of colors $N_c$, similarly to how this has been done in the conventional spectral measurements in \cite{Athenodorou:2016ebg}.
When these goals are achieved, it will be possible to check that the shape of the leading Regge trajectory agrees with the effective string theory predictions at large spins (this may require extending our analysis to $J>8$).

As we discussed, due to level broadening, the ASA stops being a useful description of the glueball spectrum at higher $N$ in all channels. Still, it will be interesting to check its predictions in some clean exclusive channels at $N>4$ with the improved precision of mass determination. A particularly interesting class of heavy states to study are the exotic $CP=-1$ $J=0$ states. At $N=5$ the ASA predicts another isolated exotic state (which can be either $0^{+-}$ or $0^{-+}$, while at $N=6$ there should be six such states (they can be either 6  $0^{+-}$s or 4 $0^{-+}$s and  2 $0^{+-}$s or evenly split between $0^{+-}$s and $0^{-+}$s).

In addition to looking at the exclusive channels the inclusive counting of states is still interesting even if different levels cannot be clearly separated as $N$ grows. Indeed, perhaps the most striking prediction of the ASA in $D=3$ YM is the absence of any massive excitations on the string worldsheet. If present, massive modes would violate the tensor square structure (\ref{tensor_square}). It is hard to test (\ref{tensor_square}) directly, when the levels start to overlap. However, the presence of massive modes would also imply a much faster growth in the number of the glueball states, compared to the ASA, which can still be tested.

On a theory side it is important to push the ASA spectrum beyond the ansatz for quantum numbers. In general, this requires a comprehensive understanding of the integrable approximation, which is lacking at the moment.
However, certain progress can be achieved, especially at large $J$, using the effective string theory technology. For instance, as clear from the glueball spectra based both on the AR and conventional basis, the mass of glueballs at the same level tend to decrease as the spin gets smaller. This fits well with the presence of an additional attractive correction to the Nambu--Goto phase shift extracted from long fluxtube data \cite{Dubovsky:2014fma,Chen:2018keo}. Indeed, the smaller $J$ glueball states are obtained by adding short wavelength  perturbation to the classical rotating rod solution, describing the leading Regge trajectory.
Additional attraction between these excitations makes their energy smaller compared to the Nambu--Goto prediction (the latter corresponds to an exactly  degenerate level). 

A further inspection of Figure~\ref{fig:Glueball-spectrum-byN-oneN5} suggests additional patterns of glueball spectrum at large $J$. Namely, it appears that at each $N$ the highest spin $(2N)^+$ state is followed by a pair of $(2N-2)^+$ and $(2N-2)^-$ states, where the latter is considerably lighter than the former, and then, perhaps, by a pair of nearly degenerate $(2N-3)^\pm$ states (at $N>2$). If correct, this pattern should be relatively straightforward to calculate within the effective string theory using the phase shift correction from  \cite{Dubovsky:2014fma,Chen:2018keo} as an input.
 
 Finally, it will be most important to extend these results to $D=4$ YM. On a theory side this is more complicated due to the presence of a massive excitation on the worldsheet of $D=4$ confining strings \cite{Athenodorou:2010cs,Dubovsky:2013gi}. Hence, it is likely that the lattice input will play an important role in achieving this goal.

\section*{Acknowledgments}
We would like to thank Guzm\'an Hern\'andez-Chifflet for helpful discussions.
This work was supported in part through the NYU IT High Performance Computing resources, services, and staff expertise.
It is also supported in part by NSF CAREER award PHY-1352119 and NSF award PHY-1915219.

\bibliographystyle{utphys}
\bibliography{dlrrefs}
\clearpage{}

\begin{table}[H]
\begin{tabular}{|c|c|c|}
\hline 
$N$ & Glueball states & \# of states\tabularnewline
\hline 
\hline 
0 & $0\otimes0=0^{++}$ & 1\tabularnewline
\hline 
1 & $\ensuremath{1\otimes1=0^{++}+0^{--}+2^{+}}$ & 4\tabularnewline
\hline 
2 & $(0+2)\otimes(0+2)=2\cdot0^{++}+0^{--}+2^{+}+2^{-}+4^{+}$ & 9\tabularnewline
\hline 
3 & $\ensuremath{(0+1+3)\otimes(0+1+3)=}$ & 25\tabularnewline
 & $\ensuremath{3\cdot0^{++}+2\cdot0^{--}+1^{+}+1^{-}+2\cdot2^{+}+2^{-}+3^{+}+3^{-}+4^{+}+4^{-}+6^{+}}$ & \tabularnewline
\hline 
4 & $(0^{P_{1}}+0^{P_{2}}+1+2+4)\otimes(0^{P_{1}}+0^{P_{2}}+1+2+4)=$ & 64\tabularnewline
 & $5\cdot0^{++}+3\cdot0^{--}+0^{P_{1}P_{2}+}+0^{P_{1}P_{2}-}+3\cdot(1^{+}+1^{-})+4\cdot2^{+}+3\cdot2^{-}+$ & \tabularnewline
 & $2\cdot(3^{+}+3^{-})+3\cdot4^{+}+2\cdot4^{-}++5^{+}+5^{-}+6^{+}+6^{-}+8^{+}$ & \tabularnewline
\hline 
\end{tabular}

\caption{Quantum numbers of 3D glueballs at first five levels, as predicted
by the ASA.\label{tab:ASA quantum numbers}}
\end{table}

\begin{table}[H]
\begin{center}
\subfloat[For the second operator in Figure \ref{fig:Set-A} (after three link smearings).
\label{tab:BadOverlaps}]{

\begin{tabular}{rrrrr}
\hline
& $J=1$ & $J=3$ & $J=5$ & $J=7$ \\ 
\hline
$J=1$ & 1.000 & 0.082 & 0.052 & 0.012 \\ 
$J=3$ & 0.082 & 1.000 & 0.020 & 0.033 \\ 
$J=5$ & 0.052 & 0.020 & 1.000 & 0.006 \\ 
$J=7$ & 0.012 & 0.033 & 0.006 & 1.000 \\ 
\hline
\end{tabular}

}\subfloat[For the operator in Figure \ref{fig:Double-operators} (after three link
smearings).\label{tab:GoodOverlaps}]{

\begin{tabular}{rrrrr}
\hline
& $J=1$ & $J=3$ & $J=5$ & $J=7$ \\ 
\hline
$J=1$ & 1.000 & 0.001 & 0.012 & 0.000 \\ 
$J=3$ & 0.001 & 1.000 & 0.000 & 0.013 \\ 
$J=5$ & 0.012 & 0.000 & 1.000 & 0.002 \\ 
$J=7$ & 0.000 & 0.013 & 0.002 & 1.000 \\ 
\hline
\end{tabular}

}\caption{Examples of absolute values of overlaps of $J^{++}$ components in
the  $E$ representation of ${\mathbb Z}_4$.}
\end{center}
\end{table}

\begin{table}[H]
\begin{tabular}{crrrrrrrrrr}
\hline
& $n=1$ & $n=2$ & $n=3$ & $n=4$ & $n=5$ & $n=6$ & $n=7$ & $n=8$ & $n=9$ & $n=10$ \\ 
\hline
$n=1$ & 0.00 & 0.00 & 0.00 & 0.02 & 0.00 & 0.01 & 0.02 & 0.01 & 0.02 & 0.02 \\ 
$n=2$ & 0.01 & 0.00 & 0.00 & 0.02 & 0.00 & 0.01 & 0.02 & 0.00 & 0.02 & 0.03 \\ 
$n=3$ & 0.02 & 0.01 & 0.00 & 0.00 & 0.01 & 0.02 & 0.03 & 0.02 & 0.01 & 0.02 \\ 
$n=4$ & 0.01 & 0.01 & 0.01 & 0.02 & 0.01 & 0.01 & 0.01 & 0.00 & 0.00 & 0.01 \\ 
$n=5$ & 0.01 & 0.02 & 0.04 & 0.01 & 0.01 & 0.01 & 0.01 & 0.00 & 0.01 & 0.03 \\ 
$n=6$ & 0.02 & 0.01 & 0.01 & 0.00 & 0.00 & 0.00 & 0.02 & 0.01 & 0.02 & 0.00 \\ 
$n=7$ & 0.00 & 0.01 & 0.01 & 0.00 & 0.00 & 0.00 & 0.05 & 0.02 & 0.02 & 0.02 \\ 
$n=8$ & 0.00 & 0.00 & 0.02 & 0.01 & 0.01 & 0.01 & 0.01 & 0.01 & 0.01 & 0.02 \\ 
$n=9$ & 0.01 & 0.01 & 0.02 & 0.01 & 0.01 & 0.01 & 0.02 & 0.03 & 0.00 & 0.01 \\ 
$n=10$ & 0.00 & 0.00 & 0.00 & 0.03 & 0.00 & 0.03 & 0.01 & 0.00 & 0.00 & 0.00 \\ 
\hline
\end{tabular}

\caption{Magnitude of overlap between $0^{++}$ (rows) and $8^{++}$ (columns)
eigenstates (at time separation 4 lattice units; before projection out of overlaps between  $0^{++}$ (rows) and $8^{++}$ states).\label{tab:BadEigenOverlaps}}
\end{table}

\pagebreak{}

\centering
\small %
\begin{longtable}[c]{ll|ll|ll}
\tabularnewline
\hline 
$J$  & \multicolumn{1}{l}{$n$} &$PC=++$ & \multicolumn{1}{l}{$PC=-+$ } & $PC=+-$ & $PC=--$ \tabularnewline
\hline
\endhead
\hline 
0 & 1  & 0.2778(5)  & 1.00(1)  & 0.897(7)  & 0.412(1) \tabularnewline
0 & 2  & 0.432(1)  & 1.01(1)  & 1.02(2)  & 0.549(2) \tabularnewline
0 & 3  & 0.562(2)  & 1.03(2)  & 1.08(2)  & 0.694(5) \tabularnewline
0 & 4  & 0.576(3)  & 1.05(2)  & 1.09(2)  & 0.709(4) \tabularnewline
0 & 5  & 0.633(4)  & 1.14(2)  & 1.13(3)  & 0.795(6) \tabularnewline
0 & 6  & 0.696(4)  & 1.15(2)  & 1.14(2)  & 0.852(7) \tabularnewline
0 & 7  & 0.710(5)  & 1.24(4)  & 1.15(3)  & 0.855(7) \tabularnewline
0 & 8  & 0.742(5)  & 1.26(3)  & 1.16(2)  & 0.86(1) \tabularnewline
0 & 9  & 0.794(7)  & 1.27(5)  & 1.21(4)  & 0.898(9) \tabularnewline
0 & 10  & 0.806(5)  & 1.30(3)  & 1.26(3)  & 0.92(1) \tabularnewline
0 & 11  & 0.814(7)  & 1.31(4)  & 1.30(3)  & 0.94(1) \tabularnewline
0 & 12  & 0.856(8)  & 1.33(3)  & 1.30(3)  & 0.95(1) \tabularnewline
0 & 13  & 0.86(1)  & 1.34(3)  & 1.33(5)  & 1.01(1) \tabularnewline
0 & 14  & 0.900(9)  & 1.34(5)  & 1.34(4)  & 1.03(2) \tabularnewline
0 & 15  & 0.92(1)  & 1.34(4)  & 1.36(5)  & 1.05(2) \tabularnewline
\hline 
4 & 1  & 0.606(3)  & 0.608(2)  & 0.708(3)  & 0.703(3) \tabularnewline
4 & 2  & 0.740(5)  & 0.730(5)  & 0.832(6)  & 0.843(5) \tabularnewline
4 & 3  & 0.815(5)  & 0.811(5)  & 0.867(8)  & 0.891(7) \tabularnewline
4 & 4  & 0.845(7)  & 0.839(6)  & 0.899(8)  & 0.93(1) \tabularnewline
4 & 5  & 0.878(8)  & 0.866(7)  & 0.918(8)  & 0.955(9) \tabularnewline
4 & 6  & 0.879(7)  & 0.870(7)  & 0.96(1)  & 1.00(1) \tabularnewline
4 & 7  & 0.94(1)  & 0.92(1)  & 1.00(1)  & 1.01(2) \tabularnewline
4 & 8  & 0.948(9)  & 0.97(1)  & 1.03(1)  & 1.02(2) \tabularnewline
4 & 9  & 0.994(9)  & 0.98(1)  & 1.04(2)  & 1.06(2) \tabularnewline
4 & 10  & 1.02(1)  & 0.98(1)  & 1.04(2)  & 1.07(2) \tabularnewline
4 & 11  & 1.02(1)  & 1.00(1)  & 1.06(1)  & 1.08(1) \tabularnewline
4 & 12  & 1.04(2)  & 1.01(2)  & 1.10(1)  & 1.11(2) \tabularnewline
4 & 13  & 1.06(2)  & 1.02(1)  & 1.10(2)  & 1.13(3) \tabularnewline
4 & 14  & 1.07(2)  & 1.03(1)  & 1.10(2)  & 1.15(3) \tabularnewline
4 & 15  & 1.07(2)  & 1.03(1)  & 1.10(2)  & 1.15(2) \tabularnewline
\hline 
8 & 1  & 0.882(6)  & 0.846(6)  & 0.958(9)  & 1.02(1) \tabularnewline
8 & 2  & 0.97(1)  & 1.03(1)  & 1.05(1)  & 1.04(2) \tabularnewline
8 & 3  & 1.04(1)  & 1.05(1)  & 1.10(2)  & 1.05(1) \tabularnewline
8 & 4  & 1.05(2)  & 1.06(1)  & 1.10(2)  & 1.11(2) \tabularnewline
8 & 5  & 1.08(2)  & 1.07(2)  & 1.12(2)  & 1.13(4) \tabularnewline
8 & 6  & 1.11(3)  & 1.11(3)  & 1.19(3)  & 1.18(2) \tabularnewline
8 & 7  & 1.17(3)  & 1.13(4)  & 1.19(3)  & 1.21(3) \tabularnewline
8 & 8  & 1.17(2)  & 1.14(2)  & 1.20(4)  & 1.22(3) \tabularnewline
8 & 9  & 1.17(2)  & 1.17(2)  & 1.21(2)  & 1.23(2) \tabularnewline
8 & 10  & 1.19(3)  & 1.18(3)  & 1.25(3)  & 1.23(5) \tabularnewline
8 & 11  & 1.19(2)  & 1.19(3)  & 1.26(6)  & 1.24(6) \tabularnewline
8 & 12  & 1.20(4)  & 1.20(4)  & 1.27(3)  & 1.31(5) \tabularnewline
8 & 13  & 1.23(4)  & 1.21(4)  & 1.27(3)  & 1.31(3) \tabularnewline
8 & 14  & 1.25(4)  & 1.25(3)  & 1.28(5)  & 1.32(5) \tabularnewline
8 & 15  & 1.25(4)  & 1.25(5)  & 1.29(4)  & 1.32(4) \tabularnewline
\hline 
\caption{Masses (in lattice units) for glueballs in $A_1$ and $A_2$ representations of ${\mathbb Z}_4$.}
\label{tab:MassesA1A2}
\end{longtable}\pagebreak{}

\begin{longtable}[c]{ll|ll|ll}
\tabularnewline
\hline 
$J$ & \multicolumn{1}{l}{$n$} & $PC=++$ & \multicolumn{1}{l}{$PC=-+$} & $PC=+-$ & $PC=--$\tabularnewline
\hline
\endhead
\hline 
1 & 1  & 0.684(4)  & 0.695(4)  & 0.654(3)  & 0.660(3) \tabularnewline
1 & 2  & 0.786(6)  & 0.796(5)  & 0.788(6)  & 0.789(6)\tabularnewline
1 & 3  & 0.827(5)  & 0.832(6)  & 0.792(4)  & 0.792(5) \tabularnewline
1 & 4  & 0.887(8)  & 0.880(8)  & 0.878(8)  & 0.875(9) \tabularnewline
1 & 5  & 0.914(9)  & 0.922(8)  & 0.881(7)  & 0.875(7) \tabularnewline
1 & 6  & 0.93(1)  & 0.93(1)  & 0.89(1)  & 0.902(7) \tabularnewline
1 & 7  & 0.93(1)  & 0.95(1)  & 0.931(9)  & 0.917(9) \tabularnewline
1 & 8  & 0.95(1)  & 0.96(1)  & 0.956(8)  & 0.94(1) \tabularnewline
1 & 9  & 1.01(1)  & 0.99(1)  & 0.99(1)  & 1.00(2) \tabularnewline
1 & 10  & 1.03(1)  & 1.01(2)  & 1.00(1)  & 1.01(1) \tabularnewline
1 & 11  & 1.04(1)  & 1.03(2)  & 1.03(2)  & 1.02(1) \tabularnewline
1 & 12  & 1.04(2)  & 1.04(2)  & 1.04(1)  & 1.02(1) \tabularnewline
1 & 13  & 1.05(2)  & 1.05(2)  & 1.05(2)  & 1.02(2) \tabularnewline
1 & 14  & 1.05(2)  & 1.06(2)  & 1.07(2)  & 1.06(2) \tabularnewline
1 & 15  & 1.08(2)  & 1.07(1)  & 1.09(2)  & 1.06(2) \tabularnewline
\hline 
3 & 1  & 0.690(3)  & 0.693(3)  & 0.685(4)  & 0.690(3) \tabularnewline
3 & 2  & 0.803(6)  & 0.802(7)  & 0.776(5)  & 0.779(5) \tabularnewline
3 & 3  & 0.803(6)  & 0.807(6)  & 0.795(6)  & 0.805(5) \tabularnewline
3 & 4  & 0.909(9)  & 0.912(9)  & 0.908(6)  & 0.895(8) \tabularnewline
3 & 5  & 0.92(1)  & 0.922(9)  & 0.918(7)  & 0.922(9) \tabularnewline
3 & 6  & 0.95(1)  & 0.952(8)  & 0.946(8)  & 0.93(1) \tabularnewline
3 & 7  & 0.95(1)  & 0.96(1)  & 0.950(9)  & 0.953(9) \tabularnewline
3 & 8  & 1.01(1)  & 1.00(1)  & 0.99(1)  & 0.97(2) \tabularnewline
3 & 9  & 1.03(1)  & 1.02(1)  & 0.99(1)  & 1.00(1) \tabularnewline
3 & 10  & 1.04(2)  & 1.04(2)  & 1.02(1)  & 1.04(2) \tabularnewline
3 & 11  & 1.04(1)  & 1.06(1)  & 1.07(2)  & 1.05(1) \tabularnewline
3 & 12  & 1.06(1)  & 1.06(2)  & 1.08(1)  & 1.06(2) \tabularnewline
3 & 13  & 1.07(2)  & 1.06(2)  & 1.10(2)  & 1.09(2) \tabularnewline
3 & 14  & 1.08(2)  & 1.08(1)  & 1.10(2)  & 1.09(2) \tabularnewline
3 & 15  & 1.09(3)  & 1.10(2)  & 1.12(2)  & 1.09(2) \tabularnewline
\hline 
5 & 1  & 0.811(5)  & 0.815(6)  & 0.804(5)  & 0.805(5) \tabularnewline
5 & 2  & 0.912(8)  & 0.919(8)  & 0.908(7)  & 0.892(7) \tabularnewline
5 & 3  & 0.91(1)  & 0.922(8)  & 0.925(7)  & 0.905(9) \tabularnewline
5 & 4  & 1.00(1)  & 1.01(1)  & 0.97(1)  & 0.98(1) \tabularnewline
5 & 5  & 1.04(2)  & 1.05(1)  & 1.04(1)  & 1.06(1) \tabularnewline
5 & 6  & 1.06(2)  & 1.07(1)  & 1.04(1)  & 1.06(1) \tabularnewline
5 & 7  & 1.07(1)  & 1.09(2)  & 1.05(1)  & 1.07(1) \tabularnewline
5 & 8  & 1.07(1)  & 1.12(1)  & 1.09(2)  & 1.09(2) \tabularnewline
5 & 9  & 1.11(2)  & 1.13(2)  & 1.12(2)  & 1.11(2) \tabularnewline
5 & 10  & 1.13(2)  & 1.15(2)  & 1.15(2)  & 1.13(2) \tabularnewline
5 & 11  & 1.13(2)  & 1.15(2)  & 1.15(2)  & 1.14(2) \tabularnewline
5 & 12  & 1.16(2)  & 1.16(2)  & 1.15(2)  & 1.15(2) \tabularnewline
5 & 13  & 1.16(2)  & 1.16(2)  & 1.16(2)  & 1.15(2) \tabularnewline
5 & 14  & 1.19(2)  & 1.17(2)  & 1.16(2)  & 1.16(3) \tabularnewline
5 & 15  & 1.20(3)  & 1.18(3)  & 1.18(2)  & 1.17(2) \tabularnewline
\hline 
7 & 1  & 0.937(9)  & 0.91(1)  & 0.917(7)  & 0.915(8) \tabularnewline
7 & 2  & 1.01(1)  & 1.04(1)  & 0.98(2)  & 1.03(1) \tabularnewline
7 & 3  & 1.04(1)  & 1.04(2)  & 1.03(1)  & 1.03(1) \tabularnewline
7 & 4  & 1.05(2)  & 1.04(2)  & 1.07(2)  & 1.09(1) \tabularnewline
7 & 5  & 1.06(2)  & 1.05(1)  & 1.09(2)  & 1.10(1) \tabularnewline
7 & 6  & 1.11(2)  & 1.10(3)  & 1.09(1)  & 1.13(1) \tabularnewline
7 & 7  & 1.14(2)  & 1.10(2)  & 1.10(2)  & 1.13(2) \tabularnewline
7 & 8  & 1.15(2)  & 1.11(2)  & 1.12(2)  & 1.14(1) \tabularnewline
7 & 9  & 1.17(3)  & 1.14(2)  & 1.13(3)  & 1.15(3) \tabularnewline
7 & 10  & 1.17(2)  & 1.15(2)  & 1.13(2)  & 1.15(2) \tabularnewline
7 & 11  & 1.18(3)  & 1.18(2)  & 1.14(2)  & 1.16(2) \tabularnewline
7 & 12  & 1.21(2)  & 1.19(4)  & 1.16(3)  & 1.18(3) \tabularnewline
7 & 13  & 1.21(3)  & 1.22(4)  & 1.19(3)  & 1.19(2) \tabularnewline
7 & 14  & 1.21(2)  & 1.22(3)  & 1.19(3)  & 1.21(3) \tabularnewline
7 & 15 & 1.22(4)  & 1.23(3)  & 1.21(2)  & 1.23(2) \tabularnewline
\hline
\caption{Masses (in lattice units) for glueballs in $E$ representation of ${\mathbb Z}_4$.}
\label{tab:MassesEbyP}
\end{longtable}\pagebreak{}

\normalsize

\begin{table}[ht]
\centering %
\begin{tabular}{rr|ll|ll}
\hline 
$J$ & \multicolumn{1}{r}{$n$} & $PC=++$ & \multicolumn{1}{l}{$PC=-+$} & $PC=+- $& $PC=--$\tabularnewline
\hline 
2 & 1  & 0.468(1)  & 0.469(1)  & 0.571(2)  & 0.570(2) \tabularnewline
2 & 2  & 0.594(3)  & 0.597(2)  & 0.696(3)  & 0.696(3) \tabularnewline
2 & 3  & 0.687(4)  & 0.679(4)  & 0.813(5)  & 0.807(6) \tabularnewline
2 & 4  & 0.710(4)  & 0.716(4)  & 0.828(6)  & 0.820(6) \tabularnewline
2 & 5  & 0.783(5)  & 0.770(7)  & 0.880(8)  & 0.883(8) \tabularnewline
2 & 6  & 0.814(6)  & 0.820(5)  & 0.90(1)  & 0.915(8) \tabularnewline
2 & 7  & 0.844(8)  & 0.837(7)  & 0.950(8)  & 0.95(1) \tabularnewline
2 & 8  & 0.845(7)  & 0.841(8)  & 0.975(9)  & 0.953(7) \tabularnewline
2 & 9  & 0.856(7)  & 0.843(8)  & 1.00(1)  & 0.98(1) \tabularnewline
2 & 10  & 0.877(9)  & 0.876(9)  & 1.00(1)  & 1.01(1) \tabularnewline
2 & 11  & 0.903(9)  & 0.901(9)  & 1.02(1)  & 1.02(2) \tabularnewline
2 & 12  & 0.94(1)  & 0.96(1)  & 1.04(2)  & 1.021(9) \tabularnewline
2 & 13  & 0.955(9)  & 0.96(1)  & 1.04(2)  & 1.04(1) \tabularnewline
2 & 14  & 0.96(1)  & 0.98(1)  & 1.05(2)  & 1.05(2) \tabularnewline
2 & 15  & 0.98(1)  & 0.98(1)  & 1.05(1)  & 1.05(2) \tabularnewline
\hline 
6 & 1  & 0.745(4)  & 0.749(4)  & 0.846(6)  & 0.839(6) \tabularnewline
6 & 2  & 0.880(7)  & 0.891(7)  & 0.91(1)  & 0.912(7) \tabularnewline
6 & 3  & 0.913(7)  & 0.906(6)  & 0.98(1)  & 0.95(1) \tabularnewline
6 & 4  & 0.940(8)  & 0.918(8)  & 1.01(1)  & 0.98(1) \tabularnewline
6 & 5  & 0.94(1)  & 0.966(9)  & 1.04(2)  & 1.05(2) \tabularnewline
6 & 6  & 0.98(1)  & 0.97(1)  & 1.06(2)  & 1.05(2) \tabularnewline
6 & 7  & 0.99(1)  & 1.00(1)  & 1.07(1)  & 1.06(1) \tabularnewline
6 & 8  & 1.01(1)  & 1.01(1)  & 1.11(2)  & 1.06(2) \tabularnewline
6 & 9  & 1.05(1)  & 1.05(1)  & 1.11(2)  & 1.08(3) \tabularnewline
6 & 10  & 1.05(1)  & 1.05(2)  & 1.13(2)  & 1.11(3) \tabularnewline
6 & 11  & 1.07(2)  & 1.08(3)  & 1.14(2)  & 1.13(2) \tabularnewline
6 & 12  & 1.09(2)  & 1.10(2)  & 1.14(2)  & 1.14(3) \tabularnewline
6 & 13  & 1.11(2)  & 1.11(3)  & 1.15(3)  & 1.14(2) \tabularnewline
6 & 14  & 1.11(3)  & 1.12(3)  & 1.17(3)  & 1.16(3) \tabularnewline
6 & 15  & 1.12(1)  & 1.12(2)  & 1.18(2)  & 1.16(3) \tabularnewline
\hline 
\end{tabular}\caption{Masses for glueballs in $A_3$ and $A_4$ representations of ${\mathbb Z}_4$.}
\label{tab:MassesA34byP}
\end{table}

\begin{table}[ht]
\centering %
\begin{tabular}{rllll}
\hline 
$n$ & $4^{+}$ & $4^{-}$ & $8^{+}$ & $8^{-}$\tabularnewline
\hline 
1  & 0.609(2)  & 0.707(2)  & 0.862(4)  & 0.989(8) \tabularnewline
2  & 0.735(3)  & 0.838(4)  & 1.004(7)  & 1.054(7) \tabularnewline
3  & 0.814(3)  & 0.876(5)  & 1.054(8)  & 1.08(1) \tabularnewline
4  & 0.840(5)  & 0.913(6)  & 1.06(1)  & 1.11(1) \tabularnewline
5  & 0.870(5)  & 0.953(7)  & 1.07(1)  & 1.18(2) \tabularnewline
6  & 0.874(4)  & 0.956(7)  & 1.14(2)  & 1.18(3) \tabularnewline
7  & 0.936(7)  & 1.009(8)  & 1.15(2)  & 1.19(1) \tabularnewline
8  & 0.959(8)  & 1.05(1)  & 1.16(2)  & 1.20(2) \tabularnewline
9  & 0.986(8)  & 1.055(7)  & 1.18(2)  & 1.24(3) \tabularnewline
10  & 1.010(9)  & 1.06(1)  & 1.18(1)  & 1.24(2) \tabularnewline
11  & 1.01(1)  & 1.09(1)  & 1.19(2)  & 1.24(2) \tabularnewline
12  & 1.013(9)  & 1.10(1)  & 1.21(3)  & 1.26(3) \tabularnewline
13  & 1.030(9)  & 1.10(1)  & 1.21(3)  & 1.28(3) \tabularnewline
14  & 1.04(1)  & 1.10(2)  & 1.24(3)  & 1.29(3) \tabularnewline
15  & 1.06(1)  & 1.12(2)  & 1.26(2)  & 1.30(2) \tabularnewline
\hline 
\end{tabular}\caption{Masses for glueballs within $A_1$ and $A_2$ representation of ${\mathbb Z}_4$ with
$J>0$ averaged over $P$.}
\label{MassesA12overP}
\end{table}

\begin{table}[ht]
\centering %
\begin{tabular}{rllll}
\hline 
$n$ & $2^{+}$ & $2^{-}$ & $6^{+}$ & $6^{-}$\tabularnewline
\hline 
1  & 0.4689(7)  & 0.571(1)  & 0.745(3)  & 0.843(4) \tabularnewline
2  & 0.598(2)  & 0.695(2)  & 0.886(5)  & 0.917(6) \tabularnewline
3  & 0.683(2)  & 0.809(4)  & 0.911(5)  & 0.971(8) \tabularnewline
4  & 0.712(3)  & 0.823(4)  & 0.930(5)  & 1.00(1) \tabularnewline
5  & 0.781(4)  & 0.880(5)  & 0.951(8)  & 1.05(1) \tabularnewline
6  & 0.819(4)  & 0.907(6)  & 0.971(8)  & 1.05(1) \tabularnewline
7  & 0.838(5)  & 0.953(5)  & 1.00(1)  & 1.06(1) \tabularnewline
8  & 0.845(6)  & 0.967(8)  & 1.00(1)  & 1.08(1) \tabularnewline
9  & 0.851(5)  & 0.986(8)  & 1.05(1)  & 1.09(2) \tabularnewline
10  & 0.876(7)  & 1.014(7)  & 1.051(9)  & 1.13(1) \tabularnewline
11  & 0.901(7)  & 1.02(1)  & 1.10(2)  & 1.14(2) \tabularnewline
12  & 0.951(8)  & 1.03(1)  & 1.10(1)  & 1.15(2) \tabularnewline
13  & 0.955(6)  & 1.045(9)  & 1.11(1)  & 1.15(1) \tabularnewline
14  & 0.970(8)  & 1.05(1)  & 1.12(1)  & 1.16(2) \tabularnewline
15  & 0.980(9)  & 1.06(1)  & 1.13(2)  & 1.18(2) \tabularnewline
\hline 
\end{tabular}\caption{Masses for glueballs within $A_3$ and $A_4$ representations of ${\mathbb Z}_4$  averaged over $P$.}
\label{MassesA34overP}
\end{table}

\begin{table}[ht]
\centering %
\begin{tabular}{rllllllll}
\hline 
$n$ & $1^{+}$ & $1^{-}$ & $3^{+}$ & $3^{-}$ & $5^{+}$ & $5^{-}$ & $7^{+}$ & $7^{-}$\tabularnewline
\hline 
1  & 0.691(3)  & 0.657(2)  & 0.690(3)  & 0.686(2)  & 0.812(4)  & 0.805(3)  & 0.925(6)  & 0.919(5) \tabularnewline
2  & 0.788(4)  & 0.784(4)  & 0.803(5)  & 0.778(4)  & 0.915(6)  & 0.898(5)  & 1.025(8)  & 1.01(1) \tabularnewline
3  & 0.830(5)  & 0.795(3)  & 0.806(4)  & 0.799(4)  & 0.921(6)  & 0.917(6)  & 1.041(9)  & 1.031(9) \tabularnewline
4  & 0.885(6)  & 0.869(6)  & 0.911(6)  & 0.900(4)  & 1.000(9)  & 0.977(8)  & 1.045(8)  & 1.08(1) \tabularnewline
5  & 0.915(6)  & 0.876(5)  & 0.921(6)  & 0.917(5)  & 1.045(9)  & 1.049(7)  & 1.08(1)  & 1.09(1) \tabularnewline
6  & 0.935(7)  & 0.892(6)  & 0.944(6)  & 0.938(5)  & 1.069(9)  & 1.06(1)  & 1.08(1)  & 1.12(1) \tabularnewline
7  & 0.939(8)  & 0.933(7)  & 0.956(8)  & 0.950(6)  & 1.09(1)  & 1.058(8)  & 1.13(1)  & 1.12(2) \tabularnewline
8  & 0.948(7)  & 0.941(6)  & 1.005(8)  & 0.98(1)  & 1.10(1)  & 1.09(1)  & 1.13(2)  & 1.13(2) \tabularnewline
9  & 0.997(9)  & 1.00(1)  & 1.030(8)  & 0.996(8)  & 1.13(1)  & 1.11(1)  & 1.16(2)  & 1.14(1) \tabularnewline
10  & 1.02(1)  & 0.997(8)  & 1.040(8)  & 1.033(9)  & 1.14(2)  & 1.15(1)  & 1.16(2)  & 1.15(2) \tabularnewline
11  & 1.03(1)  & 1.027(9)  & 1.06(1)  & 1.06(1)  & 1.15(2)  & 1.15(1)  & 1.20(1)  & 1.15(2) \tabularnewline
12  & 1.05(1)  & 1.03(1)  & 1.06(1)  & 1.07(1)  & 1.16(1)  & 1.15(1)  & 1.20(2)  & 1.20(2) \tabularnewline
13  & 1.06(1)  & 1.05(1)  & 1.06(1)  & 1.10(1)  & 1.16(2)  & 1.15(1)  & 1.21(3)  & 1.21(2) \tabularnewline
14  & 1.06(1)  & 1.06(1)  & 1.08(1)  & 1.10(1)  & 1.18(3)  & 1.15(1)  & 1.22(2)  & 1.21(2) \tabularnewline
15  & 1.06(1)  & 1.06(1)  & 1.09(1)  & 1.11(1)  & 1.19(2)  & 1.18(2)  & 1.22(2)  & 1.21(2) \tabularnewline
\hline 
\end{tabular}\caption{Masses of glueballs within $E$ representation of ${\mathbb Z}_4$ averaged
over $P$.}
\label{MassesEoverP}
\end{table}

\begin{table}[ht]
\centering %
\begin{tabular}{|rrrl|llll|llll|}
\hline 
N & $J^{PC}$ & n & Mass & N & $J^{PC}$ & n & Mass & N & $J^{PC}$ & n & Mass\tabularnewline
\hline 
0 & $0^{++}$  & 1  & 0.2778(5)  & 4 & $0^{++}$  & 8  & 0.742(5)  & \textgreater 4 & $0^{++}$  & 14  & 0.900(9) \tabularnewline
 &  &  &  & 4 & $0^{++}$  & 9  & 0.794(7)  & \textgreater 4 & $0^{+-}$  & 2  & 1.02(2) \tabularnewline
1 & $0^{++}$  & 2  & 0.432(1)  & 4 & $0^{++}$  & 10  & 0.806(5)  & \textgreater 4 & $0^{-+}$  & 1  & 1.00(1) \tabularnewline
1 & $0^{--}$  & 1  & 0.412(1)  & 4 & $0^{++}$  & 11  & 0.814(7)  & \textgreater 4 & $0^{--}$  & 8  & 0.86(1) \tabularnewline
1 & $2^{\pm+}$  & 1  & 0.4689(7)  & 4 & $0^{++}$  & 12  & 0.856(8)  & \textgreater 4 & $1^{\pm+}$  & 5  & 0.915(6) \tabularnewline
 &  &  &  & 4 & $0^{++}$  & 13  & 0.86(1)  & \textgreater 4 & $1^{\pm-}$  & 5  & 0.876(5) \tabularnewline
2 & $0^{++}$  & 3  & 0.562(2)  & 4 & $0^{+-}$  & 1  & 0.897(7)  & \textgreater 4 & $2^{\pm+}$  & 9  & 0.851(5) \tabularnewline
2 & $0^{++}$  & 4  & 0.576(3)  & 4 & $0^{--}$  & 5  & 0.795(6)  & \textgreater 4 & $2^{\pm-}$  & 6  & 0.907(6) \tabularnewline
2 & $0^{--}$  & 2  & 0.549(2)  & 4 & $0^{--}$  & 6  & 0.852(7)  & \textgreater 4 & $3^{\pm+}$  & 4  & 0.911(6) \tabularnewline
2 & $2^{\pm+}$  & 2  & 0.598(2)  & 4 & $0^{--}$  & 7  & 0.855(7)  & \textgreater 4 & $3^{\pm-}$  & 4  & 0.900(4) \tabularnewline
2 & $2^{\pm-}$  & 1  & 0.571(1)  & 4 & $1^{\pm+}$  & 2  & 0.788(4)  & \textgreater 4 & $4^{\pm+}$  & 6  & 0.874(4) \tabularnewline
2 & $4^{\pm+}$  & 1  & 0.609(2)  & 4 & $1^{\pm+}$  & 3  & 0.830(5)  & \textgreater 4 & $4^{\pm-}$  & 4  & 0.913(6) \tabularnewline
 &  &  &  & 4 & $1^{\pm+}$  & 4  & 0.885(6)  & \textgreater 4 & $5^{\pm+}$  & 2  & 0.915(6) \tabularnewline
3 & $0^{++}$  & 5  & 0.633(4)  & 4 & $1^{\pm-}$  & 2  & 0.784(4)  & \textgreater 4 & $5^{\pm-}$  & 2  & 0.898(5) \tabularnewline
3 & $0^{++}$  & 6  & 0.696(4)  & 4 & $1^{\pm-}$  & 3  & 0.795(3)  & \textgreater 4 & $6^{\pm+}$  & 3  & 0.911(5) \tabularnewline
3 & $0^{++}$  & 7  & 0.710(5)  & 4 & $1^{\pm-}$  & 4  & 0.869(6)  & \textgreater 4 & $6^{\pm-}$  & 2  & 0.917(6) \tabularnewline
3 & $0^{--}$  & 3  & 0.694(5)  & 4 & $2^{\pm+}$  & 5  & 0.781(4)  & \textgreater 4 & $7^{\pm+}$  & 1  & 0.925(6) \tabularnewline
3 & $0^{--}$  & 4  & 0.709(4)  & 4 & $2^{\pm+}$  & 6  & 0.819(4)  & \textgreater 4 & $7^{\pm-}$  & 1  & 0.919(5) \tabularnewline
3 & $1^{\pm+}$  & 1  & 0.691(3)  & 4 & $2^{\pm+}$  & 7  & 0.838(5)  & \textgreater 4 & $8^{\pm+}$  & 2  & 1.004(7) \tabularnewline
3 & $1^{\pm-}$  & 1  & 0.657(2)  & 4 & $2^{\pm+}$  & 8  & 0.845(6)  & \textgreater 4 & $8^{\pm-}$  & 1  & 0.989(8)\tabularnewline
3 & $2^{\pm+}$  & 3  & 0.683(2)  & 4 & $2^{\pm-}$  & 3  & 0.809(4)  &  &  &  & \tabularnewline
3 & $2^{\pm+}$  & 4  & 0.712(3)  & 4 & $2^{\pm-}$  & 4  & 0.823(4)  &  &  &  & \tabularnewline
3 & $2^{\pm-}$  & 2  & 0.695(2)  & 4 & $2^{\pm-}$  & 5  & 0.880(5)  &  &  &  & \tabularnewline
3 & $3^{\pm+}$  & 1  & 0.690(3)  & 4 & $3^{\pm+}$  & 2  & 0.803(5)  &  &  &  & \tabularnewline
3 & $3^{\pm-}$  & 1  & 0.686(2)  & 4 & $3^{\pm+}$  & 3  & 0.806(4)  &  &  &  & \tabularnewline
3 & $4^{\pm+}$  & 2  & 0.735(3)  & 4 & $3^{\pm-}$  & 2  & 0.778(4)  &  &  &  & \tabularnewline
3 & $4^{\pm-}$  & 1  & 0.707(2)  & 4 & $3^{\pm-}$  & 3  & 0.799(4)  &  &  &  & \tabularnewline
3 & $6^{\pm+}$  & 1  & 0.745(3)  & 4 & $4^{\pm+}$  & 3  & 0.814(3)  &  &  &  & \tabularnewline
 &  &  &  & 4 & $4^{\pm+}$  & 4  & 0.840(5)  &  &  &  & \tabularnewline
 &  &  &  & 4 & $4^{\pm+}$  & 5  & 0.870(5)  &  &  &  & \tabularnewline
 &  &  &  & 4 & $4^{\pm-}$  & 2  & 0.838(4)  &  &  &  & \tabularnewline
 &  &  &  & 4 & $4^{\pm-}$  & 3  & 0.876(5)  &  &  &  & \tabularnewline
 &  &  &  & 4 & $5^{\pm+}$  & 1  & 0.812(4)  &  &  &  & \tabularnewline
 &  &  &  & 4 & $5^{\pm-}$  & 1  & 0.805(3)  &  &  &  & \tabularnewline
 &  &  &  & 4 & $6^{\pm+}$  & 2  & 0.886(5)  &  &  &  & \tabularnewline
 &  &  &  & 4 & $6^{\pm-}$  & 1  & 0.843(4)  &  &  &  & \tabularnewline
 &  &  &  & 4 & $8^{\pm+}$  & 1  & 0.862(4)  &  &  &  & \tabularnewline
\hline 
\end{tabular}\caption{Glueball masses for AR operators
  organized by N. Masses averaged over P for $J>0$. Only first
  eigenstate in each $J^{PC}$ channel shown for $N>4$ }
\label{MassesByN}
\end{table}

\begin{table}[ht]
\centering
\small

\begin{tabular}{l|llll|llll}
\hline 
 & \multicolumn{8}{l}{Overlap with AR eigenstates }\tabularnewline
\hline 
 & \multicolumn{4}{l|}{At equal times} & \multicolumn{4}{l}{At 4 lattice steps time separation}\tabularnewline
\hline 
Mass & J=0 & J=4 & J=8 & Total & J=0 & J=4 & J=8 & Total\tabularnewline
\hline 
0.2732(4) & 0.968(5) & 0.0112(4) & 0.0046(2) & 0.968(5) & \textbf{\textcolor{blue}{0.998(5)}} & 0.021(2) & 0.025(3) & 0.999(5)\tabularnewline
0.409(2) & 0.872(5) & 0.0103(3) & 0.0118(3) & 0.872(5) & \textbf{\textcolor{green}{0.985(5)}} & 0.026(2) & 0.052(3) & 0.987(5)\tabularnewline
0.506(3) & 0.742(4) & 0.0145(3) & 0.0226(3) & 0.743(4) & \textbf{\textcolor{lime}{0.929(5)}} & 0.032(3) & 0.085(4) & 0.933(5)\tabularnewline
0.535(3) & 0.765(4) & 0.0344(3) & 0.0137(2) & 0.766(4) & \textbf{\textcolor{lime}{0.914(5)}} & 0.036(2) & 0.045(4) & 0.916(5)\tabularnewline
0.556(8) & 0.0620(5) & 0.719(4) & 0.0215(3) & 0.722(4) & 0.071(2) & \textbf{\textcolor{lime}{0.904(5)}} & 0.040(3) & 0.907(5)\tabularnewline
0.594(4) & 0.599(3) & 0.0573(4) & 0.0168(3) & 0.602(3) & \textbf{\textcolor{col3}{0.729(4)}} & 0.079(2) & 0.055(4) & 0.735(4)\tabularnewline
0.62(1) & 0.638(3) & 0.0286(3) & 0.0128(3) & 0.639(3) & \textbf{\textcolor{col3}{0.780(5)}} & 0.038(3) & 0.042(4) & 0.782(5)\tabularnewline
0.622(3) & 0.657(4) & 0.0434(3) & 0.0119(3) & 0.659(4) & \textbf{\textcolor{col3}{0.806(4)}} & 0.050(3) & 0.031(5) & 0.808(4)\tabularnewline
0.656(9) & 0.0553(5) & 0.682(4) & 0.0272(3) & 0.685(4) & 0.068(2) & \textbf{\textcolor{col3}{0.809(5)}} & 0.041(5) & 0.812(5)\tabularnewline
0.658(4) & 0.581(3) & 0.0473(4) & 0.0177(3) & 0.583(3) & \textbf{\textcolor{col4}{0.670(3)}} & 0.062(3) & 0.047(4) & 0.675(3)\tabularnewline
0.663(7) & 0.473(3) & 0.0455(4) & 0.0184(3) & 0.475(3) & \textbf{\textcolor{col4}{0.643(4)}} & 0.060(3) & 0.066(4) & 0.649(5)\tabularnewline
0.69(1) & 0.541(3) & 0.0570(5) & 0.0182(3) & 0.544(3) & \textbf{\textcolor{col4}{0.628(4)}} & 0.083(2) & 0.039(5) & 0.635(4)\tabularnewline
0.70(1) & 0.1186(7) & 0.429(2) & 0.0200(3) & 0.446(2) & 0.148(3) & \textbf{\textcolor{col4}{0.636(4)}} & 0.038(5) & 0.654(4)\tabularnewline
0.703(8) & 0.508(3) & 0.0820(5) & 0.0151(3) & 0.515(3) & \textbf{\textcolor{col4}{0.637(4)}} & 0.132(3) & 0.042(5) & 0.652(4)\tabularnewline
0.72(2) & 0.259(1) & 0.366(2) & 0.0259(3) & 0.449(2) & 0.342(3) & \textbf{\textcolor{col4}{0.452(3)}} & 0.051(5) & 0.569(4)\tabularnewline
0.722(7) & 0.345(2) & 0.279(1) & 0.0279(3) & 0.445(2) & \textbf{\textcolor{col4}{0.435(4)}} & 0.357(3) & 0.056(4) & 0.565(4)\tabularnewline
0.72(2) & 0.436(2) & 0.181(1) & 0.0229(3) & 0.472(3) &  \textbf{\textcolor{col4}{0.493(3)}} & 0.219(2) & 0.050(5) & 0.542(4)\tabularnewline
0.74(1) & 0.415(2) & 0.1467(8) & 0.0287(3) & 0.441(2) & \emph{0.484(4)} & 0.204(3) & 0.063(5) & 0.529(4)\tabularnewline
0.75(1) & 0.282(2) & 0.216(1) & 0.0262(3) & 0.356(2) & 0.354(3) & \emph{0.385(3)} & 0.066(4) & 0.527(4)\tabularnewline
0.795(8) & 0.463(3) & 0.255(1) & 0.0291(3) & 0.529(3) & \emph{0.534(4)} & 0.320(3) & 0.084(4) & 0.628(4)\tabularnewline
\hline 
\end{tabular}\caption{Masses from glueballs in $A_1$, $C=+$  representations using conventional
operators and overlaps with AR eigenstates.}
\label{tab:OldMassesA1+}
\end{table}

\begin{table}[ht]
\centering
\small

\begin{tabular}{l|llll|llll}
\hline 
 & \multicolumn{8}{l}{Overlap with AR eigenstates}\tabularnewline
\hline 
 & \multicolumn{4}{l|}{At equal times} & \multicolumn{4}{l}{At 4 lattice steps time separation}\tabularnewline
\hline 
Mass & J=0 & J=4 & J=8 & Total & J=0 & J=4 & J=8 & Total\tabularnewline
\hline 
0.575(5) & 0.0378(3) & 0.694(4) & 0.0356(4) & 0.696(4) & 0.161(5) & \textbf{\textcolor{lime}{0.873(5)}} & 0.047(4) & 0.889(5)\tabularnewline
0.661(4) & 0.0486(3) & 0.732(4) & 0.0437(4) & 0.734(4) & 0.163(6) & \textbf{\textcolor{col3}{0.852(5)}} & 0.052(3) & 0.869(5)\tabularnewline
0.71(1) & 0.0284(3) & 0.447(2) & 0.0243(3) & 0.449(2) & 0.080(6) & \textbf{\textcolor{col4}{0.659(4)}} & 0.044(5) & 0.666(4)\tabularnewline
0.73(1) & 0.0200(3) & 0.466(2) & 0.0375(3) & 0.468(2) & 0.065(6) & \textbf{\textcolor{col4}{0.587(4)}} & 0.049(4) & 0.593(4)\tabularnewline
0.77(4) & 0.0365(3) & 0.328(2) & 0.0523(4) & 0.334(2) & 0.085(5) & \textbf{\textcolor{col4}{0.522(4)}} & 0.098(5) & 0.538(5)\tabularnewline
0.78(1) & 0.0680(4) & 0.569(3) & 0.0926(6) & 0.581(3) & 0.146(4) & \emph{0.668(4)} & 0.113(4) & 0.693(4)\tabularnewline
0.80(4) & 0.0271(3) & 0.178(1) & 0.0295(3) & 0.183(1) & 0.069(6) & \emph{0.275(4)} & 0.053(5) & 0.289(4)\tabularnewline
0.80(1) & 0.0708(4) & 0.449(2) & 0.0936(6) & 0.464(2) & 0.147(6) & \emph{0.544(4)} & 0.143(4) & 0.581(5)\tabularnewline
0.81(3) & 0.0397(3) & 0.1569(9) & 0.0208(3) & 0.1632(9) & 0.128(6) & \emph{0.252(4)} & 0.047(5) & 0.287(5)\tabularnewline
0.822(8) & 0.0698(4) & 0.291(2) & 0.0333(2) & 0.301(2) & 0.161(6) & 0.401(4) & 0.056(5) & 0.436(4)\tabularnewline
0.83(4) & 0.0677(5) & 0.171(1) & 0.0674(4) & 0.196(1) & 0.162(6) & \emph{0.273(4)} & 0.138(5) & 0.346(5)\tabularnewline
0.86(2) & 0.0728(5) & 0.532(3) & 0.0575(4) & 0.540(3) & 0.107(7) & 0.664(5) & 0.090(6) & 0.679(5)\tabularnewline
0.87(3) & 0.0677(5) & 0.308(2) & 0.0626(4) & 0.322(2) & 0.137(6) & 0.396(4) & 0.095(6) & 0.430(5)\tabularnewline
0.88(3) & 0.0496(4) & 0.373(2) & 0.0854(5) & 0.386(2) & 0.086(8) & 0.480(5) & 0.158(6) & 0.512(5)\tabularnewline
0.90(2) & 0.0504(4) & 0.611(3) & 0.0956(6) & 0.621(3) & 0.110(8) & 0.803(5) & 0.119(5) & 0.820(5)\tabularnewline
0.97(2) & 0.0894(5) & 0.615(3) & 0.0807(5) & 0.626(3) & 0.167(6) & 0.755(5) & 0.117(5) & 0.782(5)\tabularnewline
0.97(3) & 0.1223(7) & 0.631(3) & 0.1410(8) & 0.658(4) & 0.279(7) & 0.775(5) & 0.183(7) & 0.844(6)\tabularnewline
0.97(2) & 0.183(1) & 0.638(3) & 0.174(1) & 0.686(4) & 0.400(8) & 0.711(5) & 0.249(5) & 0.853(7)\tabularnewline
0.98(1) & 0.1485(8) & 0.400(2) & 0.0767(4) & 0.433(2) & 0.238(9) & 0.547(5) & 0.110(6) & 0.607(6)\tabularnewline
0.99(5) & 0.216(1) & 0.504(3) & 0.0945(6) & 0.556(3) & 0.480(7) & 0.619(5) & 0.125(6) & 0.793(6)\tabularnewline
\hline 
\end{tabular}\caption{Masses from glueballs in $A_2$, $C=+$  representations using conventional
 operators and overlaps with AR eigenstates.}
\label{tab:OldMassesA2+}
\end{table}
\begin{table}[ht]
\centering
\small %
\begin{tabular}{l|llll|llll}
\hline 
 & \multicolumn{8}{l}{Overlap with AR eigenstates}\tabularnewline
\hline 
 & \multicolumn{4}{l|}{At equal time} & \multicolumn{4}{l}{At 4 lattice steps time separation}\tabularnewline
\hline 
Mass & J=0 & J=4 & J=8 & Total & J=0 & J=4 & J=8 & Total\tabularnewline
\hline 
0.64(1) & 0.0282(3) & 0.510(3) & 0.0164(3) & 0.511(3) & 0.093(4) & \textbf{\textcolor{col3}{0.758(5)}} & 0.048(5) & 0.766(5)\tabularnewline
0.735(7) & 0.0706(4) & 0.693(4) & 0.0447(3) & 0.698(4) & 0.174(6) & \textbf{\textcolor{col4}{0.818(5)}} & 0.048(5) & 0.837(5)\tabularnewline
0.79(4) & 0.191(1) & 0.318(2) & 0.0154(3) & 0.371(2) & 0.365(4) & \textbf{\textcolor{col4}{0.424(4)}} & 0.028(6) & 0.560(5)\tabularnewline
0.788(8) & 0.0384(3) & 0.377(2) & 0.0424(3) & 0.381(2) & 0.091(6) & \emph{0.466(3)} & 0.079(5) & 0.482(3)\tabularnewline
0.80(1) & 0.1090(7) & 0.304(2) & 0.0169(3) & 0.323(2) & 0.220(5) & \emph{0.413(4)} & 0.035(5) & 0.469(4)\tabularnewline
0.80(2) & 0.0189(2) & 0.1647(9) & 0.0756(5) & 0.182(1) & 0.036(5) & \emph{0.324(3)} & 0.151(5) & 0.359(4)\tabularnewline
0.81(4) & 0.0821(5) & 0.1459(8) & 0.0401(3) & 0.172(1) & 0.114(4) & 0.259(4) & 0.075(5) & 0.293(4)\tabularnewline
0.83(1) & 0.229(1) & 0.195(1) & 0.0260(3) & 0.302(2) & \textbf{\textcolor{col4}{0.342(4)}} & 0.277(5) & 0.057(6) & 0.444(4)\tabularnewline
0.88(2) & 0.0439(3) & 0.1212(6) & 0.0448(3) & 0.1364(7) & 0.079(6) & 0.215(4) & 0.113(6) & 0.255(5)\tabularnewline
0.90(3) & 0.0535(4) & 0.221(1) & 0.0417(3) & 0.231(1) & 0.128(6) & 0.324(4) & 0.071(6) & 0.355(5)\tabularnewline
0.91(1) & 0.254(1) & 0.513(3) & 0.0798(5) & 0.578(3) & 0.322(5) & 0.617(5) & 0.115(5) & 0.705(6)\tabularnewline
0.92(2) & 0.220(1) & 0.419(2) & 0.0567(4) & 0.477(3) & 0.283(5) & 0.515(5) & 0.090(5) & 0.595(5)\tabularnewline
0.93(3) & 0.1020(6) & 0.371(2) & 0.0594(4) & 0.389(2) & 0.132(6) & 0.491(4) & 0.098(6) & 0.518(4)\tabularnewline
0.94(1) & 0.0818(6) & 0.579(3) & 0.0767(5) & 0.590(3) & 0.109(6) & 0.749(5) & 0.093(6) & 0.763(5)\tabularnewline
0.97(2) & 0.328(2) & 0.438(2) & 0.0748(5) & 0.552(3) & 0.407(5) & 0.550(5) & 0.099(4) & 0.691(6)\tabularnewline
0.99(5) & 0.219(1) & 0.417(2) & 0.261(1) & 0.539(3) & 0.363(6) & 0.558(6) & 0.378(7) & 0.765(8)\tabularnewline
1.01(3) & 0.281(2) & 0.480(3) & 0.1046(7) & 0.566(3) & 0.398(6) & 0.660(6) & 0.172(7) & 0.789(6)\tabularnewline
1.01(3) & 0.299(2) & 0.379(2) & 0.1327(8) & 0.500(3) & 0.462(7) & 0.520(7) & 0.212(9) & 0.727(7)\tabularnewline
1.08(5) & 0.1505(8) & 0.486(3) & 0.0889(6) & 0.516(3) & 0.305(7) & 0.623(5) & 0.123(7) & 0.705(6)\tabularnewline
1.11(2) & 0.1263(7) & 0.354(2) & 0.0890(6) & 0.386(2) & 0.26(1) & 0.468(6) & 0.142(9) & 0.554(8)\tabularnewline
\hline 
\end{tabular}\caption{Masses from glueballs in $A_1$, $C=-$ representations using conventional
 operators and overlaps with AR  eigenstates.}
\label{tab:OldMassesA1-}
\end{table}
\begin{table}[ht]
\centering
\small %
\begin{tabular}{l|llll|llll}
\hline 
 & \multicolumn{8}{l}{Overlap with AR eigenstates}\tabularnewline
\hline 
 & \multicolumn{4}{l|}{At equal time} & \multicolumn{4}{l}{At 4 lattice steps time separation}\tabularnewline
\hline 
Mass & J=0 & J=4 & J=8 & Total & J=0 & J=4 & J=8 & Total\tabularnewline
\hline 
0.400(1)  & 0.911(5)  & 0.0122(3)  & 0.0134(3)  & 0.912(5)  & \textbf{\textcolor{green}{1.012(5)}} & 0.021(3)  & 0.069(4)  & 1.015(5) \tabularnewline
0.500(3)  & 0.763(4)  & 0.0089(3)  & 0.0257(3)  & 0.764(4)  & \textbf{\textcolor{lime}{0.952(5) }} & 0.030(3)  & 0.105(4)  & 0.959(5) \tabularnewline
0.58(1)  & 0.582(3)  & 0.0202(3)  & 0.0313(3)  & 0.583(3)  & \textbf{\textcolor{col3}{0.754(5) }} & 0.046(4)  & 0.123(4)  & 0.765(5) \tabularnewline
0.619(9)  & 0.574(3)  & 0.0824(5)  & 0.0192(3)  & 0.580(3)  & \textbf{\textcolor{col3}{0.779(5) }} & 0.108(3)  & 0.063(5)  & 0.789(5) \tabularnewline
0.651(5)  & 0.0812(5)  & 0.540(3)  & 0.0180(3)  & 0.546(3)  & 0.114(2)  & \textbf{\textcolor{col3}{0.765(5) }} & 0.040(5)  & 0.774(5) \tabularnewline
0.675(4)  & 0.583(3)  & 0.0286(3)  & 0.0288(3)  & 0.584(3)  & \textbf{\textcolor{col4}{0.698(4) }} & 0.045(3)  & 0.107(5)  & 0.708(5) \tabularnewline
0.69(1)  & 0.543(3)  & 0.0859(6)  & 0.0220(3)  & 0.550(3)  & \textbf{\textcolor{col4}{0.638(4) }} & 0.090(3)  & 0.059(5)  & 0.647(4) \tabularnewline
0.720(7)  & 0.398(2)  & 0.311(2)  & 0.0260(3)  & 0.505(3)  & \textbf{\textcolor{col4}{0.569(4) }} & 0.390(3)  & 0.059(5)  & 0.692(4) \tabularnewline
0.728(6)  & 0.329(2)  & 0.442(2)  & 0.0312(3)  & 0.552(3)  & 0.429(3)  & \textbf{\textcolor{col4}{0.558(3) }} & 0.057(5)  & 0.706(4) \tabularnewline
0.74(1)  & 0.429(2)  & 0.1483(8)  & 0.0274(3)  & 0.455(2)  & \emph{0.524(4) } & 0.195(3)  & 0.060(5)  & 0.563(4) \tabularnewline
0.76(1)  & 0.542(3)  & 0.202(1)  & 0.0221(3)  & 0.579(3)  & \emph{0.597(4)}  & 0.264(3)  & 0.045(6)  & 0.654(4) \tabularnewline
0.78(3)  & 0.211(1)  & 0.1573(9)  & 0.0203(3)  & 0.264(1)  & 0.321(4)  & \textbf{\textcolor{col4}{0.246(4) }} & 0.039(6)  & 0.407(4) \tabularnewline
0.795(6)  & 0.392(2)  & 0.350(2)  & 0.0349(3)  & 0.527(3)  & \emph{0.508(4)}  & 0.442(3)  & 0.057(6)  & 0.676(4) \tabularnewline
0.796(8)  & 0.338(2)  & 0.424(2)  & 0.0418(3)  & 0.543(3)  & 0.445(4)  & \emph{0.522(4) } & 0.062(5)  & 0.689(4) \tabularnewline
0.805(7)  & 0.164(1)  & 0.1474(8)  & 0.0182(3)  & 0.221(1)  & 0.243(4)  & \emph{0.275(4)}  & 0.045(6)  & 0.370(4) \tabularnewline
0.82(3)  & 0.301(2)  & 0.157(1)  & 0.0378(3)  & 0.342(2)  & \emph{0.382(4}) & 0.240(3)  & 0.085(5)  & 0.459(4) \tabularnewline
0.820(9)  & 0.377(2)  & 0.0890(6)  & 0.0223(3)  & 0.388(2)  & 0.459(3)  & 0.169(3)  & 0.061(6)  & 0.493(4) \tabularnewline
0.824(7)  & 0.1751(9)  & 0.1031(7)  & 0.0729(4)  & 0.216(1)  & 0.233(3)  & 0.210(4)  & 0.217(6)  & 0.382(4) \tabularnewline
0.83(1)  & 0.285(2)  & 0.1167(7)  & 0.0456(4)  & 0.312(2)  & 0.335(4)  & 0.233(5)  & 0.106(5)  & 0.421(5) \tabularnewline
0.83(3)  & 0.487(3)  & 0.0886(6)  & 0.0504(3)  & 0.497(3)  & 0.558(5)  & 0.100(5)  & 0.121(7)  & 0.579(5) \tabularnewline
\hline 
\end{tabular}\caption{Masses from glueballs in $A_2$, $C=-$ representations using conventional
 operators and overlaps with AR eigenstates.}
\label{tab:OldMassesA2-}
\end{table}
\begin{sidewaystable}[ph]
\centering
\small %
\begin{tabular}{l|lllll|lllll}
\hline 
 & \multicolumn{10}{l}{Overlap with AR eigenstates}\tabularnewline
\hline 
 & \multicolumn{5}{l|}{At equal time} & \multicolumn{5}{l}{At 4 lattice steps time separation}\tabularnewline
\hline 
Mass & J=1 & J=3 & J=5 & J=7 & Total & J=1 & J=3 & J=5 & J=7 & Total\tabularnewline
\hline 
0.638(4) & 0.276(1) & 0.465(2) & 0.0210(2) & 0.0295(2) & 0.542(3) & 0.383(2) & \textbf{\textcolor{col3}{0.685(4)}} & 0.038(3) & 0.046(2) & 0.787(4)\tabularnewline
0.653(3) & 0.395(2) & 0.226(1) & 0.0436(3) & 0.0190(3) & 0.458(2) & \textbf{\textcolor{col3}{0.595(3)}} & 0.360(2) & 0.065(2) & 0.030(3) & 0.699(4)\tabularnewline
0.714(5) & 0.358(2) & 0.547(3) & 0.0337(2) & 0.0237(2) & 0.655(4) & 0.440(3) & \textbf{\textcolor{col4}{0.650(4)}} & 0.047(2) & 0.032(2) & 0.787(4)\tabularnewline
0.72(1) & 0.271(1) & 0.189(1) & 0.0296(3) & 0.0226(1) & 0.332(2) & \textbf{\textcolor{col4}{0.438(3)}} & 0.258(2) & 0.042(3) & 0.041(2) & 0.512(3)\tabularnewline
0.717(9) & 0.525(3) & 0.431(2) & 0.0946(5) & 0.0151(2) & 0.686(4) & \textbf{\textcolor{col4}{0.676(4)}} & 0.498(3) & 0.103(2) & 0.028(3) & 0.846(5)\tabularnewline
0.754(7) & 0.207(1) & 0.307(2) & 0.0642(4) & 0.0191(2) & 0.376(2) & 0.300(2) & \textbf{\textcolor{col4}{0.422(3)}} & 0.131(2) & 0.037(3) & 0.535(4)\tabularnewline
0.77(1) & 0.276(2) & 0.299(2) & 0.0511(3) & 0.0363(2) & 0.412(2) & \emph{0.397(2)} & 0.383(3) & 0.081(2) & 0.051(2) & 0.560(3)\tabularnewline
0.774(7) & 0.1088(6) & 0.196(1) & 0.1430(8) & 0.0386(3) & 0.269(1) & 0.173(2) & 0.244(2) & \textbf{\textcolor{col4}{0.323(3)}} & 0.064(2) & 0.445(3)\tabularnewline
0.79(1) & 0.222(1) & 0.1200(7) & 0.0608(3) & 0.0296(2) & 0.261(1) & \emph{0.370(3)} & 0.171(2) & 0.133(2) & 0.048(2) & 0.431(3)\tabularnewline
0.791(8) & 0.1305(7) & 0.1443(8) & 0.0564(3) & 0.0225(2) & 0.204(1) & 0.215(2) & \emph{0.251(3)} & 0.102(2) & 0.042(3) & 0.348(3)\tabularnewline
0.82(1) & 0.221(1) & 0.1651(9) & 0.0782(5) & 0.0451(3) & 0.290(2) & 0.319(3) & 0.261(2) & 0.116(2) & 0.064(2) & 0.433(3)\tabularnewline
0.83(1) & 0.287(2) & 0.278(2) & 0.222(1) & 0.0639(4) & 0.461(2) & 0.374(3) & 0.370(3) & 0.282(2) & 0.077(2) & 0.601(4)\tabularnewline
0.84(1) & 0.290(2) & 0.261(1) & 0.298(2) & 0.0547(3) & 0.494(3) & 0.375(3) & 0.346(2) & 0.379(2) & 0.068(3) & 0.639(4)\tabularnewline
0.850(9) & 0.1179(7) & 0.1383(8) & 0.245(1) & 0.0373(2) & 0.307(2) & 0.192(2) & 0.196(2) & 0.314(2) & 0.065(2) & 0.422(3)\tabularnewline
0.86(1) & 0.435(2) & 0.207(1) & 0.308(2) & 0.0792(4) & 0.577(3) & 0.557(4) & 0.254(2) & 0.387(2) & 0.084(2) & 0.729(4)\tabularnewline
0.87(1) & 0.218(1) & 0.1571(9) & 0.190(1) & 0.0785(4) & 0.338(2) & 0.281(3) & 0.214(2) & 0.257(2) & 0.111(2) & 0.451(3)\tabularnewline
0.885(8) & 0.206(1) & 0.181(1) & 0.179(1) & 0.0844(5) & 0.339(2) & 0.278(3) & 0.241(2) & 0.235(3) & 0.128(3) & 0.455(3)\tabularnewline
0.885(6) & 0.336(2) & 0.414(2) & 0.371(2) & 0.1157(6) & 0.660(4) & 0.433(3) & 0.497(4) & 0.451(3) & 0.139(3) & 0.810(5)\tabularnewline
0.90(1) & 0.1180(6) & 0.188(1) & 0.0808(5) & 0.0467(3) & 0.240(1) & 0.178(2) & 0.262(3) & 0.137(3) & 0.068(2) & 0.352(3)\tabularnewline
0.92(1) & 0.246(1) & 0.519(3) & 0.1676(9) & 0.0900(5) & 0.605(3) & 0.339(2) & 0.681(4) & 0.203(2) & 0.125(3) & 0.797(5)\tabularnewline
\hline 
\end{tabular}\caption{Masses of glueballs in $E$  representation with $C=+$ and averaged over
$P$, using conventional operators and overlaps
with AR eigenstates.}
\label{tab:OldMassesE+}
\end{sidewaystable}
\begin{sidewaystable}[ph]
\centering
\small %
\begin{tabular}{l|lllll|lllll}
\hline 
 & \multicolumn{10}{l}{Overlap with AR eigenstates }\tabularnewline
\hline 
 & \multicolumn{5}{l|}{At equal time} & \multicolumn{5}{l}{At 4 lattice steps time separation}\tabularnewline
\hline 
Mass & J=1 & J=3 & J=5 & J=7 & Total & J=1 & J=3 & J=5 & J=7 & Total\tabularnewline
\hline 
0.617(3)  & 0.692(4)  & 0.0776(4)  & 0.0200(2)  & 0.0163(2)  & 0.697(4)  & \textbf{\textcolor{col3}{0.880(5) }} & 0.097(1)  & 0.026(2)  & 0.025(2)  & 0.886(5) \tabularnewline
0.637(4)  & 0.0889(5)  & 0.614(3)  & 0.0419(3)  & 0.0257(3)  & 0.622(3)  & 0.118(1)  & \textbf{\textcolor{col3}{0.835(5) }} & 0.060(2)  & 0.034(2)  & 0.846(5) \tabularnewline
0.693(6)  & 0.507(3)  & 0.264(1)  & 0.0416(3)  & 0.0188(2)  & 0.574(3)  & \textbf{\textcolor{col4}{0.632(3)}}  & 0.338(2)  & 0.056(2)  & 0.035(2)  & 0.720(4) \tabularnewline
0.700(4)  & 0.268(1)  & 0.416(2)  & 0.0572(4)  & 0.0273(2)  & 0.499(3)  & 0.336(2)  & \textbf{\textcolor{col4}{0.601(3) }} & 0.070(2)  & 0.036(2)  & 0.693(4) \tabularnewline
0.71(1)  & 0.324(2)  & 0.280(2)  & 0.0362(2)  & 0.0269(2)  & 0.430(2)  & \textbf{\textcolor{col4}{0.523(4}})  & 0.357(3)  & 0.065(2)  & 0.048(3)  & 0.638(4) \tabularnewline
0.719(8)  & 0.205(1)  & 0.486(3)  & 0.0446(3)  & 0.0256(2)  & 0.530(3)  & 0.301(2)  & \textbf{\textcolor{col4}{0.608(4) }} & 0.067(2)  & 0.034(3)  & 0.682(4) \tabularnewline
0.742(6)  & 0.0902(5)  & 0.0623(4)  & 0.344(2)  & 0.0385(3)  & 0.363(2)  & 0.126(1)  & 0.091(2)  & \textbf{\textcolor{col4}{0.584(4) }} & 0.067(3)  & 0.608(4) \tabularnewline
0.776(6)  & 0.475(3)  & 0.289(2)  & 0.0510(3)  & 0.0250(2)  & 0.559(3)  & \textbf{\textcolor{col4}{0.601(4) }} & 0.342(3)  & 0.072(2)  & 0.047(2)  & 0.697(4) \tabularnewline
0.78(1)  & 0.355(2)  & 0.477(3)  & 0.0568(4)  & 0.0431(3)  & 0.599(3)  & 0.445(3) & \emph{0.545(3)}  & 0.086(2)  & 0.075(2)  & 0.713(4) \tabularnewline
0.79(1)  & 0.340(2)  & 0.293(2)  & 0.1186(7)  & 0.0332(2)  & 0.466(2)  & \emph{0.468(3) } & 0.361(2)  & 0.179(2)  & 0.055(2)  & 0.620(4) \tabularnewline
0.803(5)  & 0.185(1)  & 0.244(1)  & 0.0614(3)  & 0.0270(2)  & 0.313(2)  & 0.266(2)  & \emph{0.326(2) } & 0.128(2)  & 0.041(2)  & 0.442(3) \tabularnewline
0.809(4)  & 0.272(1)  & 0.270(1)  & 0.1215(6)  & 0.0422(3)  & 0.404(2)  & \emph{0.363(2) } & 0.342(2)  & 0.161(2)  & 0.075(2)  & 0.529(3) \tabularnewline
0.815(9)  & 0.1375(8)  & 0.196(1)  & 0.0847(5)  & 0.0499(3)  & 0.259(1)  & 0.198(2)  & 0.263(2)  & 0.193(2)  & 0.097(3)  & 0.394(3) \tabularnewline
0.823(7)  & 0.179(1)  & 0.1403(8)  & 0.1165(6)  & 0.0433(3)  & 0.259(1)  & 0.269(2)  & 0.213(2)  & 0.169(2)  & 0.089(3)  & 0.393(3) \tabularnewline
0.840(7)  & 0.276(1)  & 0.1510(8)  & 0.362(2)  & 0.0674(4)  & 0.484(3)  & 0.373(3)  & 0.208(3)  & 0.437(3) & 0.113(2)  & 0.622(4) \tabularnewline
0.85(1)  & 0.232(1)  & 0.353(2)  & 0.1076(6)  & 0.0787(5)  & 0.443(2)  & 0.311(2)  & 0.424(3)  & 0.141(2)  & 0.142(3)  & 0.563(3) \tabularnewline
0.85(1)  & 0.306(2)  & 0.456(2)  & 0.1433(8)  & 0.0541(3)  & 0.571(3)  & 0.382(3)  & 0.563(3)  & 0.172(2)  & 0.086(3)  & 0.707(4) \tabularnewline
0.850(7)  & 0.213(1)  & 0.1505(8)  & 0.541(3)  & 0.0913(5)  & 0.607(3)  & 0.280(2)  & 0.197(2)  & 0.640(4)  & 0.148(2)  & 0.741(4) \tabularnewline
0.857(9)  & 0.416(2)  & 0.320(2)  & 0.234(1)  & 0.0622(4)  & 0.578(3)  & 0.532(4)  & 0.390(3)  & 0.285(2)  & 0.081(3)  & 0.723(4) \tabularnewline
0.863(9)  & 0.217(1)  & 0.195(1)  & 0.0771(4)  & 0.0387(3)  & 0.304(2)  & 0.298(3)  & 0.263(3)  & 0.099(2)  & 0.065(2)  & 0.414(3) \tabularnewline
\hline 
\end{tabular}\caption{Masses from glueballs in $E$  representations with $C=-$ and averaged over
parity, using conventional  operators and overlaps
with AR eigenstates.}
\label{tab:OldMassesE-}
\end{sidewaystable}

\begin{table}[ht]
\centering
\small %
\begin{tabular}{lllllll}
\hline 
 & \multicolumn{6}{l}{Overlap with AR eigenstates }\tabularnewline
\hline 
 & \multicolumn{3}{l}{At equal time} & \multicolumn{3}{l}{At 4 lattice steps time separation}\tabularnewline
\hline 
Mass & J=2 & J=6 & Total & J=2 & J=6 & Total\tabularnewline
\hline 
0.453(1)  & 0.893(5)  & 0.0188(2)  & 0.893(5)  & \textbf{\textcolor{green}{0.988(5) }} & 0.027(2)  & 0.988(5) \tabularnewline
0.548(2)  & 0.757(4)  & 0.0158(2)  & 0.757(4)  & \textbf{\textcolor{lime}{0.902(5) }} & 0.029(2)  & 0.902(5) \tabularnewline
0.623(2)  & 0.626(3)  & 0.0229(2)  & 0.626(3)  & \textbf{\textcolor{col3}{0.805(4) }} & 0.042(2)  & 0.806(4) \tabularnewline
0.628(7)  & 0.619(3)  & 0.0381(3)  & 0.620(3)  & \textbf{\textcolor{col3}{0.747(4) }} & 0.059(2)  & 0.749(4) \tabularnewline
0.685(7)  & 0.537(3)  & 0.0434(3)  & 0.539(3)  & \textbf{\textcolor{col4}{0.637(4)}}\textbf{ } & 0.069(2)  & 0.640(4) \tabularnewline
0.69(1)  & 0.383(2)  & 0.1587(9)  & 0.414(2)  & 0.562(4)  & \textbf{\textcolor{col3}{0.273(3)}}\textbf{ } & 0.625(4) \tabularnewline
0.696(4)  & 0.486(3)  & 0.1449(8)  & 0.507(3)  & \textbf{\textcolor{col4}{0.620(4) }} & 0.231(2)  & 0.661(4) \tabularnewline
0.715(4)  & 0.358(2)  & 0.208(1)  & 0.413(2)  & \textbf{\textcolor{col4}{0.431(3) }} & 0.334(3)  & 0.545(4) \tabularnewline
0.74(1)  & 0.410(2)  & 0.1049(6)  & 0.423(2)  & \textbf{\textcolor{col4}{0.550(4) }} & 0.134(3)  & 0.566(4) \tabularnewline
0.76(1)  & 0.288(2)  & 0.505(3)  & 0.581(3)  & 0.346(3)  & \textbf{\textcolor{col4}{0.598(3) }} & 0.691(4) \tabularnewline
0.779(6)  & 0.425(2)  & 0.338(2)  & 0.543(3)  & \emph{0.510(3)}  & 0.406(3) & 0.652(4) \tabularnewline
0.78(1)  & 0.500(3)  & 0.0783(4)  & 0.506(3)  & \emph{0.583(4)}  & 0.119(2)  & 0.595(4) \tabularnewline
0.794(9)  & 0.1609(9)  & 0.315(2)  & 0.354(2)  & 0.215(3) & \emph{0.461(3)}  & 0.509(4) \tabularnewline
0.808(7)  & 0.334(2)  & 0.1033(6)  & 0.349(2)  & \emph{0.439(3)} & 0.170(3)  & 0.471(3) \tabularnewline
0.82(1)  & 0.588(3)  & 0.1156(7)  & 0.599(3)  & 0.703(4)  & 0.162(3)  & 0.721(4) \tabularnewline
0.82(1)  & 0.210(1)  & 0.1490(8)  & 0.257(1)  & 0.316(3)  & 0.267(3)  & 0.414(3) \tabularnewline
0.822(8)  & 0.254(1)  & 0.1165(6)  & 0.279(2)  & 0.331(3)  & 0.231(3)  & 0.403(4) \tabularnewline
0.84(1)  & 0.349(2)  & 0.1199(6)  & 0.369(2)  & 0.442(3)  & 0.203(3)  & 0.486(3) \tabularnewline
0.85(1)  & 0.220(1)  & 0.1259(7)  & 0.254(1)  & 0.300(3)  & 0.203(3)  & 0.362(3) \tabularnewline
0.87(1)  & 0.451(2)  & 0.0925(5)  & 0.461(2)  & 0.540(4)  & 0.152(3)  & 0.561(4) \tabularnewline
\hline 
\end{tabular}\caption{Masses from glueballs in $A_3$ and $A_4$  representations with
$C=+$ and averaged over $P$, using conventional  operators and overlaps with
AR eigenstates}
\label{tab:OldMassesA34+}
\end{table}
\begin{table}[ht]
\centering
\small %
\begin{tabular}{lllllll}
\hline 
 & \multicolumn{6}{l}{Overlap with AR eigenstates}\tabularnewline
\hline 
 & \multicolumn{3}{l}{At equal time} & \multicolumn{3}{l}{At 4 lattice steps time separation}\tabularnewline
\hline 
Mass & J=2 & J=6 & Total & J=2 & J=6 & Total\tabularnewline
\hline 
0.540(2)  & 0.782(4)  & 0.0201(2)  & 0.782(4)  & \textbf{\textcolor{lime}{0.956(5) }} & 0.048(2)  & 0.958(5) \tabularnewline
0.621(3)  & 0.662(4)  & 0.0314(3)  & 0.662(4)  & \textbf{\textcolor{col3}{0.838(5)}}\textbf{ } & 0.053(2)  & 0.839(5) \tabularnewline
0.698(4)  & 0.569(3)  & 0.0260(3)  & 0.569(3)  & \textbf{\textcolor{col4}{0.702(4) }} & 0.045(2)  & 0.703(4) \tabularnewline
0.712(8)  & 0.387(2)  & 0.0362(3)  & 0.389(2)  & \textbf{\textcolor{col4}{0.557(3) }} & 0.074(2)  & 0.562(3) \tabularnewline
0.75(2)  & 0.182(1)  & 0.238(1)  & 0.300(2)  & 0.235(2)  & \textbf{\textcolor{col4}{0.457(4)}}  & 0.514(4) \tabularnewline
0.755(6)  & 0.492(3)  & 0.0773(5)  & 0.498(3)  & \textbf{\textcolor{col4}{0.637(4) }} & 0.156(3)  & 0.656(4) \tabularnewline
0.781(6)  & 0.302(2)  & 0.0460(3)  & 0.306(2)  & \emph{0.445(3)}  & 0.086(2)  & 0.453(3) \tabularnewline
0.791(8)  & 0.246(1)  & 0.0780(5)  & 0.258(1)  & \emph{0.371(3)}  & 0.148(3)  & 0.399(3) \tabularnewline
0.79(2)  & 0.361(2)  & 0.0772(5)  & 0.369(2)  & \emph{0.513(3)}  & 0.142(3)  & 0.532(3) \tabularnewline
0.79(1)  & 0.295(2)  & 0.1452(8)  & 0.329(2)  & \emph{0.388(3) } & 0.246(2)  & 0.459(3) \tabularnewline
0.84(1)  & 0.1569(9)  & 0.1333(7)  & 0.206(1)  & 0.234(3)  & 0.223(3)  & 0.323(3) \tabularnewline
0.85(1)  & 0.325(2)  & 0.182(1)  & 0.372(2)  & 0.409(4)  & 0.286(3)  & 0.499(4) \tabularnewline
0.85(2)  & 0.386(2)  & 0.323(2)  & 0.504(3)  & 0.476(4)  & 0.447(3)  & 0.653(4) \tabularnewline
0.86(2)  & 0.348(2)  & 0.1386(8)  & 0.375(2)  & 0.483(4)  & 0.212(3)  & 0.527(4) \tabularnewline
0.87(1)  & 0.275(1)  & 0.282(2)  & 0.394(2)  & 0.374(3)  & 0.371(3)  & 0.527(4) \tabularnewline
0.87(2)  & 0.249(1)  & 0.256(1)  & 0.357(2)  & 0.349(3)  & 0.340(4)  & 0.487(4) \tabularnewline
0.90(1)  & 0.268(1)  & 0.221(1)  & 0.347(2)  & 0.377(4)  & 0.298(3)  & 0.481(4) \tabularnewline
0.91(2)  & 0.199(1)  & 0.319(2)  & 0.376(2)  & 0.282(4)  & 0.433(4)  & 0.517(4) \tabularnewline
0.947(9)  & 0.393(2)  & 0.1103(6)  & 0.408(2)  & 0.550(4)  & 0.160(3)  & 0.573(4) \tabularnewline
0.96(2)  & 0.320(2)  & 0.1651(9)  & 0.360(2)  & 0.443(4)  & 0.247(4)  & 0.507(4) \tabularnewline
\hline 
\end{tabular}\caption{Masses from glueballs in $A_3$ and $A_4$  representations with
$C=-$ averaged over $P$, using conventional  operators and overlaps with
AR  eigenstates}
\label{tab:OldMassesA34-}
\end{table}

\begin{table}[htb]
\begin{center}
\begin{tabular}{|cc|cc|c|c|}\hline
\multicolumn{6}{|c|}{ $M/M_{0^{++}}$} \\ \hline
$J^{PC}$ & $n$ & \multicolumn{2}{|c|}{$\beta=86$} & $\beta=74$ & $\beta=63$ \\
   &    & $70^280$   & $50^280$  & $60^268$ & $50^256$ \\ \hline
$2^{-+},6^{-+}$ & 1 &1.669(14)  & 1.670(11)  & 1.674(8) & 1.667(9) \\
        & 2 & 2.045(10) & 2.032(11) & 2.025(11) & 1.961(15) \\
        & 3 & 2.295(14) & 2.321(17) & 2.301(15) & 2.308(26) \\
        & 4 & 2.334(26) & 2.353(26) & 2.378(20) & 2.372(13) \\
        & 5 & 2.600(17) & 2.545(33) & 2.575(15) & 2.582(28) \\
        & 6 & 2.604(36) & 2.641(17) & 2.605(20) & 2.595(14) \\ \hline
$4^{-+}$  & 1 & 2.154(23) & 2.154(14) & 2.164(12) & 2.055(48) \\
        & 2 & 2.392(26) & 2.417(20) & 2.452(18) & 2.460(33) \\ \hline 
$1^{\pm +},3^{\pm +}$  & 1 & 2.356(16) & 2.359(33) & 2.375(15) & 2.326(23) \\
        & 2 & 2.404(29) & 2.417(36) & 2.405(18) & 2.366(19) \\
        & 3 & 2.626(16) & 2.580(39) & 2.619(33) & 2.552(92) \\ \hline
$1^{\pm -},3^{\pm -}$  &  1 & 2.289(13) & 2.260(33) & 2.296(18) & 2.269(23) \\
        & 2 & 2.366(13) & 2.356(14) & 2.386(12) & 2.310(23) \\
        & 3 & 2.543(36) & 2.593(23) & 2.597(20) & 2.547(33) \\ \hline\hline
$aM_{0^{++}}$  & & 0.3115(11) & 0.3120(11) & 0.3650(12) & 0.4350(15) \\ 
$a\surd\sigma$  & & 0.07378(8) & -- & 0.08637(10) & 0.10252(11) \\ \hline
\end{tabular}
\caption{Some mass ratios using conventional operators. To test for
  finite volume and lattice spacing corrections. Last two rows list
  the mass gap and the square root of the string tension.}
\label{table_mM}
\end{center}
\end{table}

\clearpage{}

\end{document}